\documentclass{emulateapj}

\shorttitle{MIR and Radio properties of SDSS AGN}
\shortauthors{Rosario et al.}

\newcommand{\kms}{km s$^{-1}$}
\newcommand{\ergs}{erg s$^{-1}$}
\newcommand{\whz}{W Hz$^{-1}$}
\newcommand{\mics}{$\mu$m}
\newcommand{\hb}{H$\beta$}
\newcommand{\ha}{H$\alpha$}
\newcommand{\othree}{[\ion{O}{3}]$\lambda 5007$}
\newcommand{\ntwo}{[\ion{N}{2}]$\lambda 6584$}
\newcommand{\lmir}{L$_{12}$}
\newcommand{\lfir}{L$_{22}$}

\newcommand{\lrad}{L$_{1.4}$}
\newcommand{\lothree}{L$_{[OIII]}$}
\newcommand{\smass}{M$_*$}
\newcommand{\msun}{M$_{\odot}$}
\newcommand{\lsun}{L$_{\odot}$}
\newcommand{\dfour}{$D_{n}4000$}

\begin{document}

\title{The Mid-infrared Emission of Narrow-Line Active Galactic Nuclei: Star-Formation, Nuclear Activity and two populations revealed by WISE}

\author{David J. Rosario, Leonard Burtscher, Richard Davies, Reinhard Genzel, Dieter Lutz, Linda J. Tacconi}

\affil{Max Planck Institute for Extraterrestrial Physics, Postfach 1312, 85741 Garching, Germany}

\begin{abstract}

We explore the nature of the long-wavelength mid-infrared (MIR) emission of a sample of 13000 local Type II (narrow-line)
Active Galactic Nuclei (AGNs) from the Sloan Digital Sky Survey (SDSS) using 
12 and 22 \mics\ photometry from the WISE all-sky survey. In combination
with FIRST 1.4 GHz photometry, we show that AGNs divide into two relatively distinct populations or ``branches" 
in the plane of MIR and radio luminosity. Seyfert galaxies lie almost exclusively on a MIR-bright branch (Branch A), while
low-ionization nuclear emission line galaxies (LINERs) are split evenly into Branch A and the MIR-faint Branch B. 
We devise various tests to constrain the processes that define the branches, including a comparison to the properties
of pure star-forming (SF) inactive galaxies on the MIR-Radio plane. We demonstrate that the total MIR emission of
objects on Branch A, including most Seyfert galaxies, is governed primarily by host star-formation,
with $\approx$15\% of the 22 \mics\ luminosity coming from AGN-heated dust. This implies that on-going dusty
star-formation is a general property of Seyfert host galaxies. We show that the 12 \mics\ broad-band 
luminosity of AGNs on Branch A is suppressed with respect to star-forming galaxies, possibly due to the 
destruction of PAHs or deeper 10 \mics\ Si absorption in AGNs. We uncover a correlation between the MIR luminosity
and \othree\ luminosity in AGNs. This suggests a relationship between the SFR and nuclear luminosity in the AGN
population, but we caution on the importance of selection effects inherent to such AGN-dominated emission-line
galaxies in driving such a correlation. We highlight the MIR-radio plane as a useful tool in comparative studies of
SF and nuclear activity in AGN.

\end{abstract}


\section{Introduction}


Through their enormous and concentrated energetic output, supermassive black holes (SMBHs) 
are expected to play an important role in the evolution of their host galaxies. As yet elusive is
a detailed understanding of the interplay between accretion onto the black hole and
their final output. Recently, patterns exhibited by stellar mass black holes
have been shown to extend to their supermassive cousins, allowing various relationships to be proposed between the accretion rate, 
radiative efficiency and mechanical output via relativistic outflows \cite[e.g.][]{maccarone03, merloni03, churazov05,trump11}.
At low specific accretion rates (Eddington ratios $\lambda_{E} \lesssim 10^{-2}$), the flow onto the SMBH
is believed to be radiatively inefficient, possibly advecting much of its thermal energy through the event horizon.
A substantial fraction of the accreted mass is channeled into radio-emitting jets,
likely mediated by the magnetic field of the SMBH. At higher $\lambda_{E}$,
the accretion flow settles into a radiatively efficient thin disk, which produces a profuse X-ray and 
extreme ultra-violet (EUV) radiation field at the expense of the relativistic outflow. The bright high-energy
spectrum of such an AGN can strongly ionize gas out to kiloparsecs, resulting in extended AGN emission line regions.

In most AGNs, the direct EUV is inaccessible due to its high optical depth to gas and dust from the 
vicinity of the SMBH and on galaxy scales. Much of the absorbed EUV is reprocessed to
the mid-infrared (MIR) from AGN-heated dust, at hundreds of K, located 
in the putative pc-scale ``torus". Indeed, AGN spectral energy distributions (SEDs) 
show clear evidence for excess hot dust in the MIR \cite[e.g.][]{sanders89, netzer07, wu09}. 
Therefore, the relative output of an AGN between the 
MIR and radio wavelengths may serve as a tracer of the accretion mode of the growing SMBH. 

There is, however, an important complication to this simple argument. The emission of galaxies in the MIR is
also influenced by other processes, most importantly the heating of very small dust grains 
by star-formation (SF) and evolved stellar populations \citep[e.g.][]{rowan-robinson89, desert90, draine01,groves12},
as well as the excitation of the complex of bands from Polycyclic Aromatic Hydrocarbons (PAHs) \citep[e.g.][]{draine07}.
Prior to an investigation of the accretion properties of AGNs, an account must be made for the 
MIR emission from these other processes. Such an approach also affords a worthwhile by-product. The
thermal infrared is a valuable tracer of star-formation and is relatively free of the effects of extinction or optical
depth saturation that affects the UV or optical emission lines \citep[e.g.][]{calzetti07}. In massive, metal-rich 
galaxies, such as AGN hosts, most of the emission from young stars is reprocessed by dust in molecular clouds
and emitted in the IR. Therefore, a detailed study of the MIR in AGNs can provide a handle on both
the accretion properties of the nuclear source and its relationship to SF in the host galaxy.

Theoretical and empirical arguments support a connection between SF and nuclear activity in AGN hosts, either by
direct synchronization between a starburst and the fueling of the nucleus \citep{sanders88,norman88,storchi01, davies07}, 
or indirectly mediated by the availability of cold gas needed for luminous AGN activity \citep{heller94, rosario13}. 
In the era of the Infrared Space Observatory (ISO) and the Spitzer space telescope, much work has 
examined the interplay between AGN and SF through the use of MIR spectroscopy \citep{genzel00, verma05, soifer08}, employing
fine structure emission lines of various ionized species \citep[e.g.][]{genzel98,sturm02,melendez08,tommasin10,diamond-stanic12}, 
PAH features and their relative strengths \citep[e.g.][]{schweitzer06, shi07,odowd09,diamond-stanic10,lamassa12}, and  
continuum luminosities and MIR colors \citep[e.g.][]{wu09,lamassa10}. These studies have greatly enhanced
our understanding of the MIR phenomenology and physics relevant to the AGN-SF connection, firming up
such results as the increased abundance of AGN signatures in IR-luminous galaxies \citep[e.g.][]{genzel98, nardini08},
the anticorrelation of PAH strength with AGN prominence \citep[e.g.][]{clavel00,odowd09, lamassa10} and the
greater depth of the 10 \mics\ Si absorption feature in Type II AGNs \citep[e.g.][]{shi06, hao07}. However, as discussed in Section 2.2.2,
most of these studies were limited to either very nearby or fairly luminous systems, selected based on the capabilities
of the early Infrared Astronomical Satellite (IRAS) \citep{soifer87, rush93}. 

A recent huge leap forward comes from the surveys from the Wide-field Infrared Survey Explorer (WISE), 
with more than two orders of magnitude deeper MIR sensitivity than IRAS over the whole sky, 
enabling a greater dynamic range and better photometry of sources over a wider range of redshifts. 
In this paper, we use MIR and radio photometry from the WISE and FIRST surveys to assess the
dominant source of the total, galaxy-integrated MIR emission in local ($z<0.15$) narrow emission line-selected AGNs from the SDSS.
The data and sample properties are laid out in Section 2. In Section 3, we discuss the diagnostics
of the MIR-radio luminosity plane and outline two distinct populations of AGNs, which we discuss in the context
of AGN ionization classes. We perform various tests using ancillary measurements from the SDSS and AKARI surveys
(Section 4) and then synthesize our findings towards an understanding of the nature of the two populations (Section 5).

Throughout this work, we assume a $\Lambda$-CDM Concordance cosmology with H$_{0} = 73$ \kms Mpc$^{-1}$ and
$\Omega_{\Lambda} = 0.7$. 

\section{Sample Selection and Data}


\subsection{SDSS Emission-Line AGNs}

Emission line-selected AGNs were drawn from the 
\anchor{http://www.mpa-garching.mpg.de/SDSS/DR7/}{MPA-JHU database of spectral measurements}, 
based on the Sloan Digital Sky Survey (SDSS) Data Release 7 \citep[DR7,][]{sdssdr7}. This database is restricted to
objects with galaxy-like spectra and narrow Balmer lines, specifically excluding broad-line (Type I) AGNs. Therefore,
our sample consists almost exclusively of narrow-line (Type II) AGNs. From the entire catalog
of 818333 unique galaxies, we chose all objects satisfying the following criteria:

\begin{itemize}
\item  A high-quality spectroscopic redshift in the range $0.02 < z < 0.15$.
\item  S/N$>3$ in the emission lines of \othree, \hb, \ha\ and \ntwo, as well as a S/N$>2$ in the weaker [\ion{S}{2}$]\lambda 6720$ line doublet.
\item  A location in the standard `BPT' diagram of \othree/\hb\ vs.~\ntwo/\ha\ \citep{bpt81, vo87} 
above the curve which separates objects with AGN-dominated ionization from composite and SF galaxies \citep{kewley01, kewley06}.
These are objects in the Seyfert and LINER domains of the BPT diagram.
\item Equivalent width of \ha$>3$\AA\ to ensure that weak LINERs are excluded from the sample. Lines in systems
with weaker \ha\ can arise primarily in shocks or through the UV field from evolved stars, 
rather than by a nuclear source \citep{cidfernandes11}.
\end{itemize}

From this subset of 13339 AGNs, we separated high-ionization AGN (Seyferts) from
LINERs using the [S II]$\lambda 6720$/\ha\ criterion of \cite{kewley06}, Eqns.~7 \& 12. The use of a simple cut in 
\othree/\hb\ ratio of 3.0 to separate high and low ionization AGNs leads to considerable
mixing at the boundaries of these populations in the standard BPT diagram. Hence, we adopted
the use of the fainter [S II] line as an additional selection criterion, even though it tends to reject
intrinsically lower luminosity AGNs with faint emission line fluxes.

We also make use of host stellar mass (\smass) and 4000\AA\ break (\dfour) measurements from the MPA-JHU database.
The masses were derived from fits to the five-band SDSS photometry of the hosts, following the methodology
of \citet{salim07}. The procedure for \dfour\ measurements is described in
\cite{kauffmann03a}. 

\subsubsection{Selection effects inherent to SDSS emission line-selected AGNs}

AGNs selected through BPT-based criteria from the SDSS spectroscopic database give us
by far the largest census of local nuclear activity. While a remarkable resource for AGN studies,
a finer understanding of the relationship between AGN activity and the properties of host galaxies requires
an appreciation of the biases inherent to the selection of such AGN samples. 

Standard BPT criteria divide emission-line galaxies into three populations: AGN-dominated, 
SF-dominated and so-called `composite' systems. The last category includes objects
that show line ratios intermediate to SF- and AGN-dominated systems, and contain a large number of AGNs
with substantial SF in their inner regions \citep[e.g.][]{kauffmann03, juneau11}. 
AGN-dominated systems, those which lie above the curve from \cite{kewley01},
have their line emission from within the SDSS aperture largely arising in AGN ionized gas.
While cleanly separating out AGNs, this criterion additionally selects against objects with strong star-formation in the central
aperture. AGN-dominated systems will have, on average, a lower SFR within the SDSS aperture
than purely SF or composite systems.

The SDSS aperture (3" is diameter) covers between $0.6$ and $3.8$ kpc (projected) in galactic radius across our
working redshift range of 0.02--0.15. This may be compared to the typical effective radii of massive
galaxies, which range from 2--8 kpc \citep{trujillo06}. In general, the SDSS aperture covers less than half
of a galaxy's light. The typical size of the Narrow-Line Region (NLR), in which most of the 
AGN-ionized emission originates, is $\sim$ kpc for bright Seyfert galaxies \citep{bennert06}. Therefore, almost 
all of the emission from the AGN will be sampled by the SDSS aperture. As the physical radius subtended by the SDSS 
aperture increases with redshift, a growing fraction of an average galaxy is sampled by the SDSS spectrum. 
While, at low redshifts, AGN-dominated systems may still have considerable star-formation outside the region
probed by the SDSS aperture, at higher redshifts, as more of the galaxy falls within the aperture, AGN-dominated
systems become increasingly special objects, containing either fairly luminous AGNs, galaxies with weak SF or
objects where most of the SF is either obscured or lies on the outskirts of the galaxy.

Consider an AGN host galaxy with a certain nuclear luminosity $L_{AGN}$ and total SFR. In general, this
SFR is spread over the host galaxy, with a fraction $f_{in}$ that lies within the SDSS fiber aperture. 
The emission line contribution purely from HII regions in the SDSS spectrum is broadly set by SFR and $f_{in}$ (we marginalize
over other factors such as host metallicity and morphology for this heuristic argument). If $L_{AGN} > L_{AGN,c}$, a
certain critical value, the emission line spectrum of the AGN-ionized gas will be brighter than that from HII regions
and the galaxy will enter the AGN-dominated region of the BPT diagram and satisfy our selection. $L_{AGN,c}$ is
proportional to the SFR and $f_{in}$ -- as $f_{in} \times$SFR increases, $L_{AGN,c}$ also increases. Therefore,
among strongly SF galaxies or galaxies at higher redshifts (where the SDSS fiber covers a larger part of the host galaxy),
AGNs have to be more luminous to enter the AGN-dominated part of the BPT diagram, leaving a larger fraction of
the active population in the composite region of the diagram.This effect will also
force a correlation between $L_{AGN}$ and SFR among line-selected AGNs, 
even if one is not physically present.  

\begin{figure}[t]
\figurenum{1}
\label{sed_plots}
\centering
\includegraphics[width=\columnwidth]{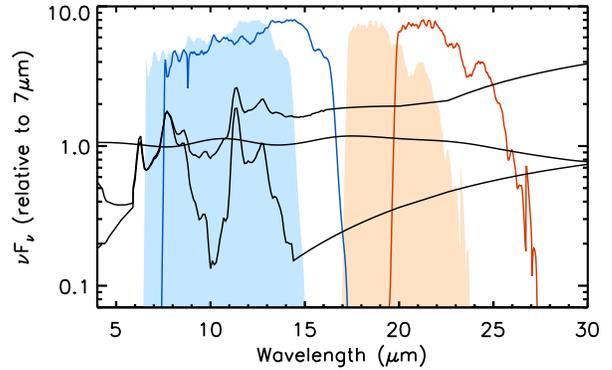}
\caption[]
{An illustration of the WISE W3 and W4 bands and the features of star-forming galaxy and AGN SEDs that
are sampled by these bands. Three rest-frame SEDs are shown, all normalized to unity at 7 \mics. The two SEDs
which show PAH features are SF templates from the \cite{ce01} library. The SED with the higher
PAH EW corresponds to a total IR luminosity of $10^{10}$ \lsun, while that with the lower PAH EW 
has a total IR luminosity of $10^{12}$ \lsun. The relatively featureless AGN SED is the mean template
from \citep{mor12}. The filter functions in the W3 and W4 WISE bands are shown as blue and red curves
respectively. The open curves correspond to $z=0$, while the filled curves show the wavelengths sampled
by the filters at $z=0.15$, the upper redshift limit of the sample studied in this work. }
\end{figure}

\subsection{WISE All-Sky Survey}

The Wide-field Infrared Survey Explorer \citep[WISE][]{wright10} has mapped the entire sky in four MIR bands at
3.4 \mics, 4.6 \mics, 12 \mics\ and 22 \mics. A public catalog is available consisting of
photometry from the all-sky survey atlas of point-like and extended sources detected with a S/N$>5$ 
in at least one of the four WISE bands. The catalog is not uniform, since the depth of the WISE imaging 
varies considerably across the sky. The astrometry accuracy of the catalog, tied to the 2MASS coordinate system, 
is better than 200 mas. 

We crossmatched the SDSS-selected AGN with the WISE all-sky survey catalog available from the IPAC/IRSA service, using a simple
cone search with a tolerance of 2". More than 98\% had a counterpart in the WISE survey, almost 75\% of which also
had S/N$>2$ detections in both 12 and 22 \mics\ bands. 

We rely on the photometry performed by the WISE all-sky survey pipeline, details of which may be found 
in the \anchor{http://wise2.ipac.caltech.edu/docs/release/allsky/expsup/}{Data Release Supplement}.
The pipeline provides several different photometric measurements for sources in all four WISE bands. The primary photometry
is performed by PSF decomposition, assuming that pipeline targets are
composed of blends of 1--2 point sources (the `profile-fitting' or PRO photometry). A maximum-likelihood model 
of the source plane is simultaneously fit in all four bands for contiguous batches of sources.
In addition, various flavors of circular and elliptical aperture photometry are also provided by the survey pipeline.

\begin{figure}[t]
\figurenum{2}
\label{iras_comp}
\centering
\includegraphics[width=\columnwidth]{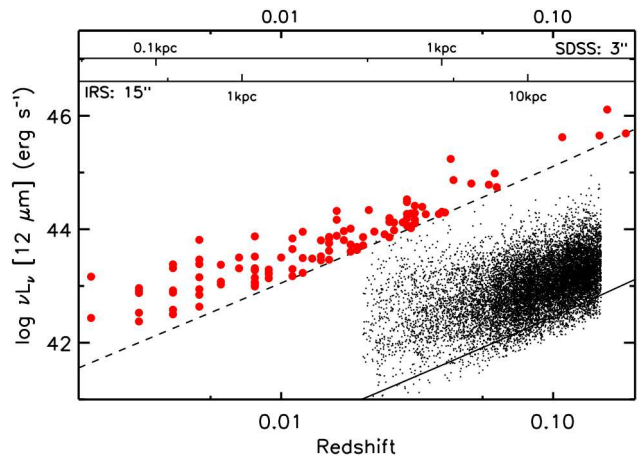}
\caption[]
{Observed-frame 12 \mics\ luminosity vs.~ redshift of SDSS emission-line selected AGNs (small black points) and
Seyferts from the IRAS extended 12 \mics\ sample with Spitzer/IRS spectra from \citet{wu09} (large red points).
The dashed line shows the limit set by the 0.22 Jy cut used to select the the extended 12 \mics\ sample. The solid
line shows a nominal limit in the W3 band from the WISE all-sky survey: the actual depth of the survey varies
across the sky leading to scatter below the line. Scale bars at the top mark the projected physical radii in kpc
spanned by an example circularized IRS aperture size of 15" and the SDSS fiber aperture of 3" diameter.
}
\end{figure}

In this work, we only employ photometry in the two longer wavelength bands. 
The FWHM of the WISE PSFs  are $\approx 6\farcs8$ in the 12 \mics\ (W3) band
and $\approx 11\farcs8$ in the 22 \mics\ (W4) band. At the lowest redshifts, galaxies could
be extended in the WISE images, even beyond the large 22 \mics\ PSF. The profile-fit $\chi^2$ in a band
serves as a way to identify if the image of a source in that band is likely to be extended. The WISE pipeline
uses a value of reduced $\chi^{2} > 3$ to flag a source as extended. We find that, of the roughly 12440 sources
in our `working sample' (those with better than 2$\sigma$ detections in either W3 or W4), 
only 264 have a reduced $\chi^2 > 3$ in either the W3 or W4 bands: 80\%\ of these
lie at $z<0.05$ and 95\%\ are associated with an extended counterpart in the 2MASS all-sky near-IR survey. 
The vast majority of the sources in our sample are unresolved in the W3 and W4 bands of WISE and for these
we adopt the PRO magnitudes as the best estimates of their photometry. 
For the resolved sources with 2MASS counterparts, the WISE pipeline generates aperture photometry
in ellipses matched and scaled to the sizes and shapes of the corresponding 2MASS sources, which we adopt
as the best estimates of their magnitudes. The W3 magnitudes of extended sources
have been boosted by a small additive factor of 0.44 mag to account for imperfect background subtraction
and other effects shown to influence the aperture photometry (see WISE All-Sky Survey 
Explanatory Supplement Sec VI.3.e). We exclude the remaining 32 extended sources with no 2MASS counterparts
from our sample. 

\begin{figure*}[t]
\figurenum{3}
\label{lmir_lradio_total}
\centering
\includegraphics[width=\textwidth]{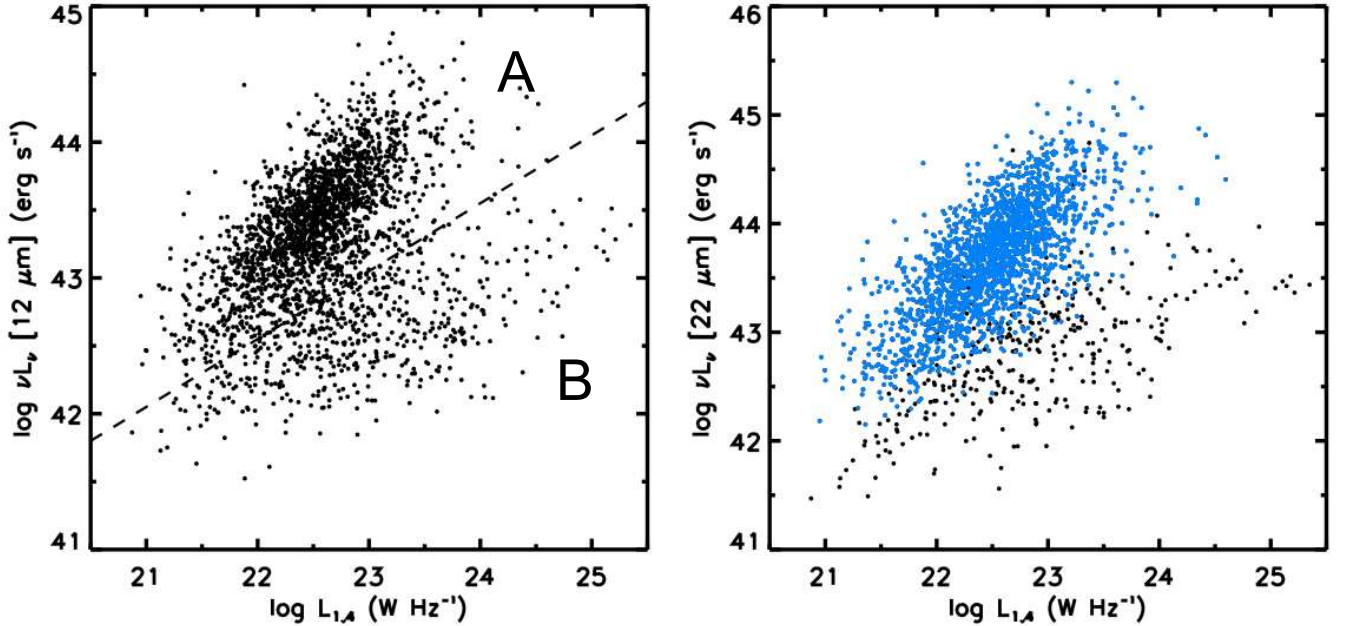}
\caption{{\bf Left:} Rest-frame 12 \mics\ luminosity (\lmir) vs.~1.4 GHz luminosity (\lrad) of 
SDSS emission-line selected AGNs in the redshift range 
$0.02<z<0.15$. Two populations or ``branches" of objects may be discerned 
which describe different slopes in the correlation between \lmir\ and \lrad. These have been marked A and B
and a dashed line is plotted which roughly divides between the branches.
{\bf Right:} Rest-frame 22 \mics\ luminosity (\lfir) vs. 1.4 GHz luminosity (\lrad) of SDSS emission-line selected AGNs in the redshift range 
$0.02<z<0.15$. Based on the division into branches from the left panel, 
objects are plotted with differently colored points: blue for objects
on Branch A and black for objects on Branch B. The division based on \lmir\ translates very cleanly into different branches
in \lfir\ vs. \lrad, highlighting the continuity between photometry in the PAH-sensitive W3 band and the continuum-dominated W4 band. 
}
\end{figure*}

\subsubsection{MIR features probed by the WISE bands}

In Figure \ref{sed_plots}, we examine the parts of the IR SEDs sampled by the two WISE bands in both 
SF and pure AGN torus-dominated cases. Two representative SF galaxy templates from the
\citet{ce01} library are plotted, corresponding to total IR (8--1000 \mics) luminosities 
of $10^{10}$ \lsun\
(a moderately SF galaxy) and $10^{12}$ \lsun\ (a local Ultra-luminous IR galaxy [ULIRG]).
For an torus-dominated SED, we show the template of \citet{mor12}, an average from 
MIR observations of local Type I AGNs spanning a wide range in luminosity. 

At all redshifts in this study, the W3 band covers the complex of emission bands at $8{\rm -}13$ \mics\ produced by 
Polycyclic Aromatic Hydrocarbons (PAHs) as well as the broad silicate absorption feature at 10 \mics. 
The W4 band, on the other hand, is free of PAHs and serves as a good measure 
of the pure dust continuum emission, which, in star-forming galaxies, comes from warm dust in
the high temperature tail of the typical distribution of grains. Due to the systematic variation
of PAH equivalent widths (EWs) with total IR luminosity \citep{ce01,dale01}, the W3-W4 colors of SF galaxies are
bluer in low luminosity IR galaxies (i.e., those with low levels of SF) than in IR luminous systems, such as ULIRGs. 

Compared to SF templates, pure AGN templates are flat and fairly featureless in the $10-30$ \mics\ wavelength band
\citep{netzer07, mullaney11a, mor12}. Theoretical studies of AGN-heated dust ``tori" suggest a broad
range of peak wavelengths around 15--30 \mics\ depending on the structure of the dust and the intrinsic AGN spectrum
\citep{hoenig06, fritz06, nenkova08}. Real AGNs, however, frequently show PAH features in their spectra,
from SF in their host galaxies. Significant silicate absorption is also found in their SEDs, especially
among Type IIs 
\citep[e.g.][]{deo07,goulding12}.
The W3-W4 color of a typical AGN lies within the range shown by SF galaxies - a simple color criterion involving 
these bands cannot easily distinguish between star-formation and AGN dominated systems. 

\subsubsection{Comparison to IRAS extended 12 \mics\ AGN sample}

In Figure \ref{iras_comp}, we plot observed-frame 12 \mics\ luminosities against redshift of SDSS AGNs with 
WISE W3 band photometry (small black points) along with a subset of AGNs from the 
IRAS-based extended 12 \mics\ sample of galaxies (large red points). The latter come from the
compilation of \citet{wu09} and constitute a large fraction of local Seyferts that have been studied spectroscopically
with the Spitzer/IRS instrument \citep{buchanan06, wu09, tommasin10, lamassa10}. The selection of the extended 12 \mics\
sample of AGNs was set by the depths of the IRAS Faint Source catalog v2 \citep{rush93} and all have 12 \mics\ fluxes 
$f_{12,IRAS} \geq 0.22$ Jy. In comparison, the WISE all-sky survey reaches down to 0.5 mJy (2$\sigma$) at 12 \mics.
Therefore, the sources typically studied using detailed IRS spectroscopy are either much nearer or much more luminous
that the vast majority of the sources in our sample. From Figure \ref{iras_comp}, we see that the typical redshift of
AGNs from the 12 \mics\ sample is $z\sim0.02$, whereas most of our AGNs lie at $z\sim0.1$. Since the scales covered by
nominal IRS apertures ($\approx 15"$) are similar to those covered by the WISE PSF, most earlier
spectroscopic studies probed AGN hosts on radii of $\lesssim$ few kpc --  just the circum-nuclear regions --  
while most of our AGNs are photometered over the entire galaxy. In addition, our objects
cover a much larger swathe of MIR luminosity than existing samples over the redshifts where the two samples overlap,
and are not restricted to the IR-luminous systems detected by IRAS. These differences should be borne in mind when
comparing our findings to results in the contemporary literature.

\begin{figure*}[t]
\figurenum{4}
\label{lmir_lradio_separate}
\centering
\includegraphics[width=\textwidth]{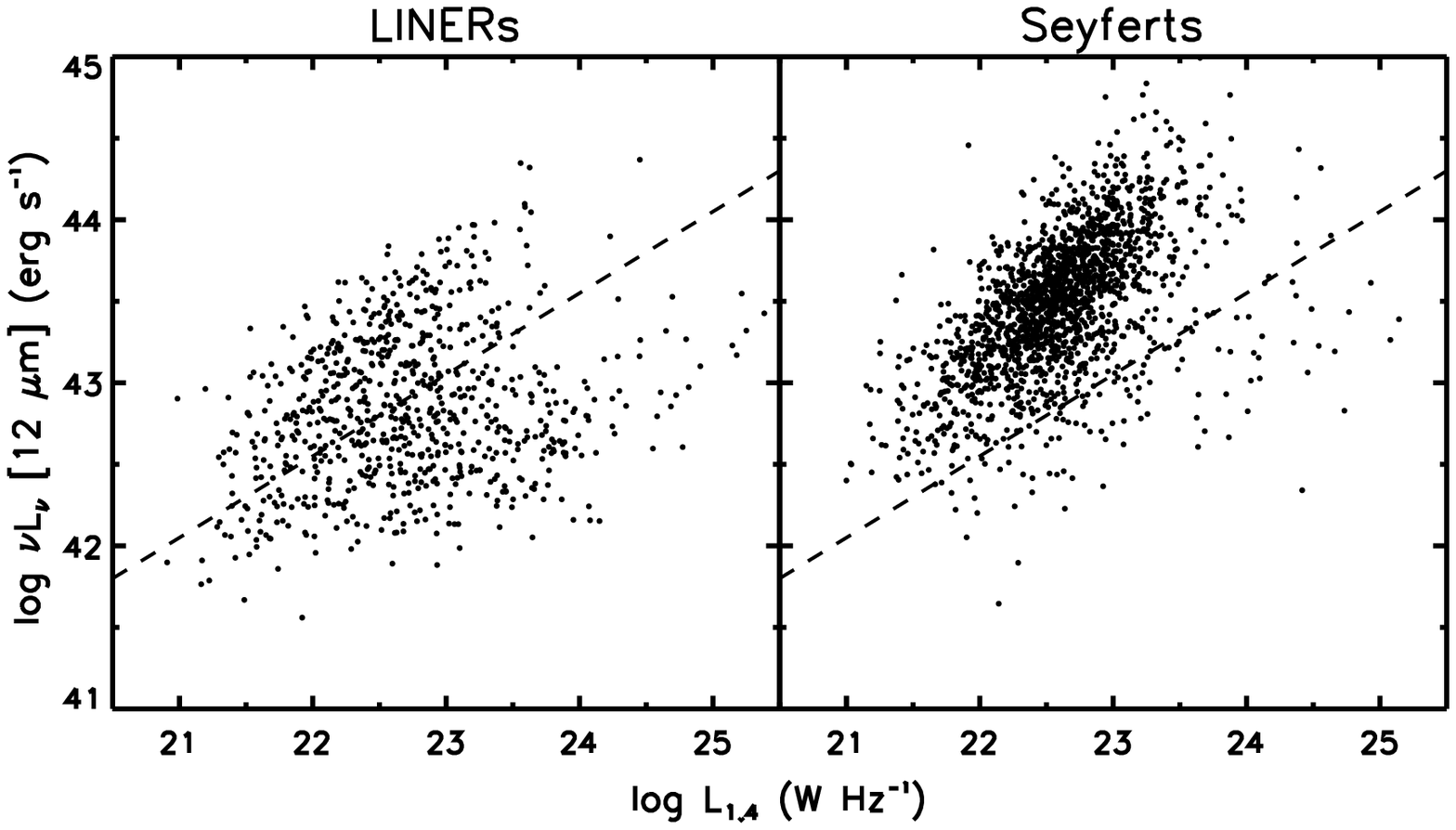}
\caption{Rest-frame 12 \mics\ luminosity (\lmir) vs. 1.4 GHz luminosity (\lrad) of SDSS emission-line selected AGNs in the redshift range 
$0.02<z<0.15$. The dashed line which divides the two branches from Figure \ref{lmir_lradio_total} is also plotted.
LINERs and Seyferts are shown in the left and right panels respectively. The two classes
of AGNs distribute very differently in this diagram. While LINERs occupy both branches, 
Seyferts are almost completely on Branch A.}
\end{figure*}

\subsection{FIRST 20 cm Radio Survey}


For a radio survey complementary in many ways to the SDSS, we turn to the VLA-FIRST
20 cm (1.4 GHz) survey  \citep{becker95}. The current public catalog combines observations of about 10,000 deg$^2$ of sky
coincident with the SDSS. The resolution of the FIRST survey is $\approx 5$" with an astrometric accuracy of 50 mas. The integrated
flux of FIRST sources is measured using two-dimensional gaussian fits to the source profile on the survey maps. For a small
fraction of sources which are very extended and jetted, those corresponding to classical radio galaxies, a gaussian profile
fit is an inadequate representation of the source structure and the FIRST photometry of such sources may be in error.
Since the vast majority of the sources in our sample are at low power (\lrad$<10^{23}$ \whz), the source sizes are likely to be
close to or below the resolution of the survey and the fluxes will be accurate.

We crossmatched the SDSS-selected AGN with the FIRST source catalog using a search tolerance of 2". This yielded
FIRST counterparts for 23\% of LINERs and 19\% of Seyferts, about $\times2$ higher than the
SDSS counterpart fraction of FIRST sources at the $r<18$ magnitude limit of the SDSS spectroscopic survey \citep{devries07}.
The difference is due to our stringent emission line selection, which picks out more luminous systems. Relaxing the $H\alpha$
EW requirement, for e.g., yields radio detection rates of $\approx 10$\%.
Due to the relative shallowness of FIRST, only a subdominant fraction of the AGN population can be probed completely by this study. 
In the rest of this work, unless otherwise stated, we use the term AGNs, LINERs or Seyferts
to refer to the subpopulations detected in both WISE and FIRST surveys.

\subsection{AKARI All-Sky survey}

The far-infrared (FIR) is dominated by radiation from cold dust with temperatures $< 100$K. In galaxies, this component
of the dust emission spectrum is produced almost exclusively in star-forming regions or from diffuse cirrus. 
In FIR luminous galaxies, the component from star-formation is paramount and, even in fairly luminous AGNs, the FIR luminosity
can be taken as a relatively clean measure of the SFR \citep[e.g.][]{netzer07, rosario12}.

In Section 3.2, we investigate the relationship between SF and the MIR luminosity of AGNs. For a sample
of bonafide star-forming AGNs to aid in this study, we identify a subset of FIR-bright AGNs 
from the \anchor{http://give website}{AKARI/FIS all-sky survey Bright Source Catalog}. We crossmatch our AGNs
with the catalog using a search tolerance of 15''. The large tolerance is warranted by the greater 
positional uncertainty of the FIR catalog due to the broad AKARI PSF at 90 \mics\ ($\approx 50''$ FWHM). 
703 sources were identified with reliable 90 \mics\ photometry (FQUAL90 = 3).
The 0.55 Jy detection limit of the AKARI/FIS catalog corresponds to a 90 \mics\ luminosity limit of $\approx 10^{9}$ \lsun\
at $z=0.02$ and $\approx 10^{10.9}$ \lsun\ at $z=0.15$, picking out moderately luminous star-forming galaxies at all redshifts.

\section{Two Populations of AGNs: The MIR-Radio plane}

\begin{figure*}[t]
\figurenum{5}
\label{lmir_withz}
\centering
\includegraphics[width=\textwidth]{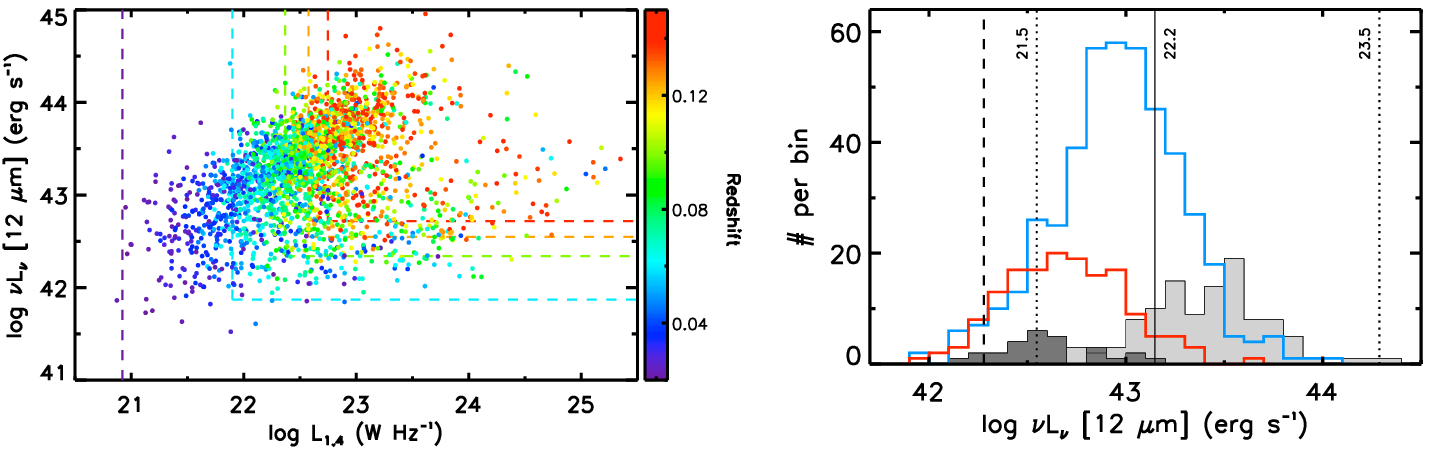}
\caption[MIR-Radio with redshift]
{{\bf Left:} Rest-frame 12 \mics\ luminosity (\lmir) vs.~1.4 GHz luminosity (\lrad) of SDSS emission-line selected AGNs
colored by redshift, according to the colorbar at right. The dashed lines show the typical luminosity limit of 
the WISE and FIRST surveys at a set of illustrative redshifts, also colored according to the same colorbar. 
Sources on Branch A are considerably brighter than the \lmir\ limits at their corresponding redshifts and the correlation
on the diagram from objects on this Branch is seen consistently at all redshifts. On the other hand,
sources on Branch B lie close to the \lmir\ limits at all redshifts. The weak correlation seen for
Branch B is driven mostly by redshift biases, in that the most luminous sources are seen only at
the highest redshifts in our sample.
{\bf Right: } \lmir\ distributions of AGNs in the narrow redshift interval $0.080<z<0.085$. The dashed vertical
line shows the typical \lmir\ limit probed by the WISE survey at these redshifts. Three vertical lines mark 
the \lmir\ values of the ridgeline of Branch A at $\log$\lrad$=$21.5 (left dotted), 22.2 (solid) 
and 23.5 (right dotted) W Hz$^{-1}$, spanning the range of radio luminosity covered by 
Branch A over the entire AGN sample. The shaded histograms
show sources that are detected in the FIRST survey. Sources on
Branch A (light gray histogram) and Branch B (dark gray histogram) show different distributions of
\lmir, as expected from their locations on the MIR-Radio plane. The colored open histograms show
sources that lie below the FIRST detection limit. The \lmir\ distribution of radio-undetected Seyferts 
(blue histogram) is consistent with most of them lying on an extrapolation of Branch A below the FIRST
detection limit. The \lmir\ distribution of radio-undetected LINERs (red histogram) is consistent with
a substantial fraction on Branch B, in parallel with radio-detected LINERs.
}
\end{figure*}

In this section, we demonstrate that AGNs divide into two relatively distinct populations in the plane of MIR and radio
luminosity. Seyferts and LINERs distinguish themselves by differentiating into these two populations in different 
proportions. We examine the effects of survey limits and whether this dichotomy is preserved even among radio-undetected AGNs. 

The rest-frame monochromatic MIR luminosities at 12 and 22 \mics\ used here are derived from the 
WISE W3 and W4 magnitudes, applying a k-correction based on a \citet{ce01} template
of a star-forming galaxy with a total IR luminosity of $10^{11}$ L$_{\odot}$. 
The k-correction is small, amounting to 0.03 mag at 12 \mics\  and 0.4 mag at 22 \mics\ for galaxies at $z=0.15$. 
The full range of IR templates from \cite{ce01} gives a maximum variation in the k-correction of $\pm 0.2$ 
magnitudes. A flat radio spectrum per unit energy was assumed, 
requiring no additional k-correction to get rest-frame 1.4 GHz luminosities. Since the radio output
of a sizable fraction of AGNs is governed by star-formation (Section 5.2.3), true spectra indices may vary as high as 
$0.5$. However, the choice of spectral index makes no significant difference to our results.

In the left panel of Figure \ref{lmir_lradio_total}, we plot the rest-frame 12 \mics\ monochromatic luminosity (\lmir) 
against the rest-frame 1.4 GHz luminosity (\lrad) of SDSS/WISE/FIRST AGNs. 
They divide quite clearly into two fairly distinct populations in the \lmir--\lrad\ plane. We use the term 
`branches' to describe the two populations and will refer to them as such in the rest of the paper. 
Objects on Branch A have a high \lmir/\lrad\ ratio and delineate a steep trend between 
12 \mics\ luminosity and radio luminosity. On Branch B, sources show a lower \lmir/\lrad\ ratio and 
a shallow dependence between \lmir\ and \lrad. The dashed line in the Figure roughly separates
the two branches and serves as a guide to the eye.

The two branches persist in the plot of \lfir\ vs.~\lrad\ (right panel of Figure \ref{lmir_lradio_total}), but the lower branch is less populated. 
This is because of the shallower depth of the WISE survey in the W4 band: IR faint objects are not significantly detected in W4,
which preferentially affects the density of objects in Branch B. Nevertheless, this diagram shares many features in 
common with the left panel, implying a close continuity between the emission of AGNs in both MIR bands. 
Indeed, if one divides sources into two branches based on their position in the left panel, 
they remain in two branches in the right panel as well (as shown with different colored points).

In Figure \ref{lmir_lradio_separate}, we show the \lmir--\lrad\ plane separately for LINERs and Seyferts. Remarkably, despite being
divided purely based on optical emission line criteria, the two AGN classes show strikingly different behavior in this diagram.
Seyferts cluster tightly and typically lie at \lmir\ $> 10^{43}$ \ergs, almost exclusively along Branch A 
($\approx 6$\% of Seyferts lie on Branch B). On the other hand, LINERs divide roughly equally between both branches.
LINERs on Branch A are slightly less luminous in \lmir\ than Seyferts of the same \lrad, tending to lie closer to the dividing line.

The division of AGNs into two branches based on MIR and radio properties
is preserved across all redshifts in our sample (left panel of Figure \ref{lmir_withz}), though the most luminous
sources, both in the MIR and the radio, are found at the higher redshifts simply because 
of the larger volumes probed. The characteristic flux
limits of the WISE and FIRST surveys correspond to an increasing \lmir\ and \lrad\ luminosity limit 
with redshift (dashed vertical and horizontal lines). Sources on Branch A are luminous enough in the MIR
that they lie well above the luminosity limit at all redshifts.  The normalization and slope of Branch A in the \lmir--\lrad\ 
plane does not appear to change with redshift, suggesting a physical basis for this Branch which remains constant to $z=0.15$. 
Sources in Branch B typically lie just around the \lmir\ limit - this suggests that the weak slope to this Branch is not
real, but is driven mostly by Malmquist bias.

\begin{figure*}[t]
\figurenum{6}
\label{lir_lradio_sfcomp}
\centering
\includegraphics[width=\textwidth]{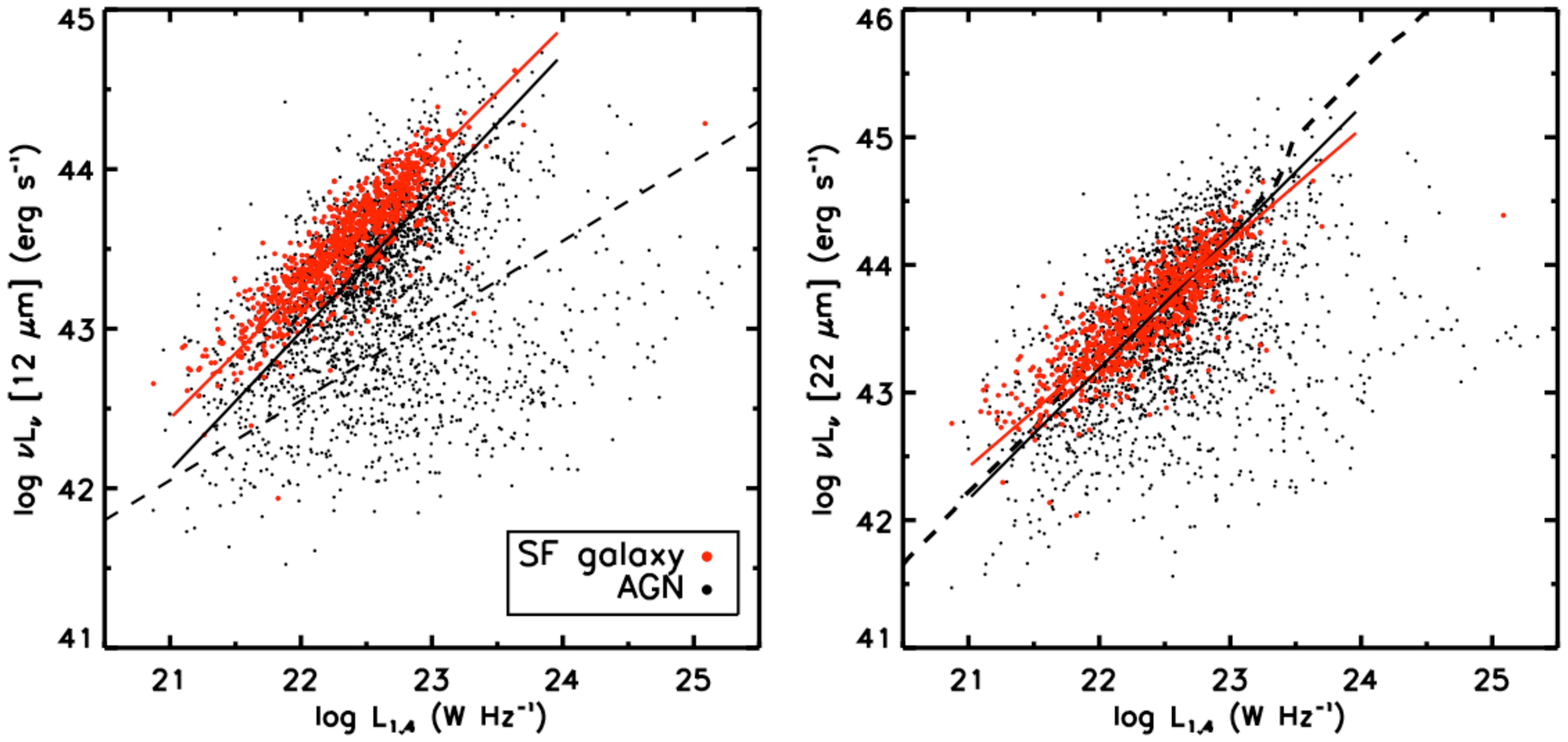}
\caption{{\bf Left:} Rest-frame 12 \mics\ luminosity (\lmir) vs. 1.4 GHz luminosity (\lrad) of SDSS selected AGNs (black points)
and a control sample of inactive star-forming galaxies (red points) in the redshift range 
$0.02<z<0.15$. The dashed line is the same as in Figure \ref{lmir_lradio_total}. Also shown are regression
lines from fits to the inactive star-forming galaxies (red line) and AGNs on Branch A (black line) that are matched to inactive galaxies.
The \lmir-\lrad\ trend for SF galaxies is offset from that of the AGNs on Branch A.
{\bf Right:} Rest-frame 22 \mics\ luminosity (\lfir) vs. 1.4 GHz luminosity (\lrad) of  SDSS selected AGNs (black points)
and a control sample of inactive star-forming galaxies (red points) in the redshift range 
$0.02<z<0.15$. Also show are regression lines from fits to the inactive star-forming 
galaxies (red line) and AGNs on Branch A (black line) that are matched to inactive galaxies.
The thick dashed line shows the FIR-radio correlation of SF galaxies from \cite{sargent10}, 
extrapolated to 22 \mics\ using SF templates from \cite{ce01}. 
Both SF galaxies and AGNs on Branch A lie along this correlation, and scatter consistently about it.
}
\end{figure*}

As discussed before, the relative shallowness of the FIRST survey ultimately limits the fraction of
AGNs that can be classified into branches, ranging from 40\% at $z=0.02$ to 15\% at $z=0.1$. In contrast,
the deep WISE photometry allows us to estimate MIR luminosities for most SDSS AGNs, since more than 90\% of the 
parent sample are detected in the W3 band. Could the \lmir\ distribution of sources undetected in FIRST 
give us an indication as to their likely location on the MIR-Radio plane? 
Is the coherence of Branch A seen at low redshifts preserved among AGNs at high redshifts that lie below the
FIRST flux limits?

To overcome the complexities of redshift-dependent limits, we consider a narrow redshift slice at $0.080<z<0.085$, 
roughly at the middle of the full range of redshifts. In the right panel of Figure \ref{lmir_withz}, we compare the \lmir\ distributions of
the radio-detected and -undetected SDSS/WISE AGNs and attempt to understand their differences. 
One may view these histograms as a projection of the MIR-Radio plane of Figure \ref{lmir_lradio_total}
onto the Y-axis for sources in the small redshift interval.  
For the 17\% of the sources detected in FIRST, we plot distributions of \lmir\ split by Branch, as shown by the grey histograms.
The \lmir\ of FIRST-detected sources is bimodal, consistent with the patterns in the MIR-Radio plane. 
Of the remaining sources, those detected in the W3 band but undetected by FIRST comprise 
77\% of the full SDSS sample at these redshifts. These are plotted as the open colored 
histograms in the Figure, with Seyferts shown in blue and LINERs shown in red. Our hypothesis is that 
a substantial fraction of LINERs should be on an extension of Branch B below the radio limit, while
Seyferts should mostly lie on Branch A. Their \lmir\ distributions should reflect this and this is indeed what
we observe.

FIRST-undetected LINERs show an \lmir\ distribution that clusters close to the MIR limit (dashed line) and
has a similar shape to that of FIRST-detected objects on Branch B, though with a tail to more luminous sources.
This is very consistent with the observed distribution of radio-detected LINERs: a substantial fraction of objects on
Branch B, with a small number on Branch A leading to the luminous tail. In contrast, the radio-undetected 
Seyferts show a distribution peaking roughly between the histograms of the two branches.
Most Seyferts have \lmir\ much higher than the MIR luminosity limit, consistent with them lying on an extension of
Branch A below the radio limit. 

We can illustrate this more quantitatively in the following manner. Since the ridgeline
of Branch A has a non-zero slope in the MIR-Radio plane, sources with higher \lrad\ correspond to higher
\lmir, with some scatter given by the width of Branch A. For a representative set of \lrad\ above, at and below the FIRST
detection limit at $z=0.08$, we calculate corresponding \lmir\ from the Branch A ridgeline (see Figure \ref{lir_lradio_sfcomp})
and plot them as vertical solid and dotted lines. The range of \lmir\ spanned by the dotted lines is equivalent to the range
spanned by the entirety of Branch A galaxies over our entire redshift range. It encompasses the Branch A histogram at this
redshift (by construction) as well as the peak of the Seyferts. 
The solid line, corresponding to the radio detection limit, delineates a transition 
between the radio-detected sources of Branch A and the peak of the radio-undetected Seyferts, as one would
expect if a substantial fraction of these Seyferts lay on an extension of Branch A.


\section{The Nature of the Two Branches}

\begin{figure*}[t]
\figurenum{7}
\label{ir_radio_ratio}
\centering
\includegraphics[width=\textwidth]{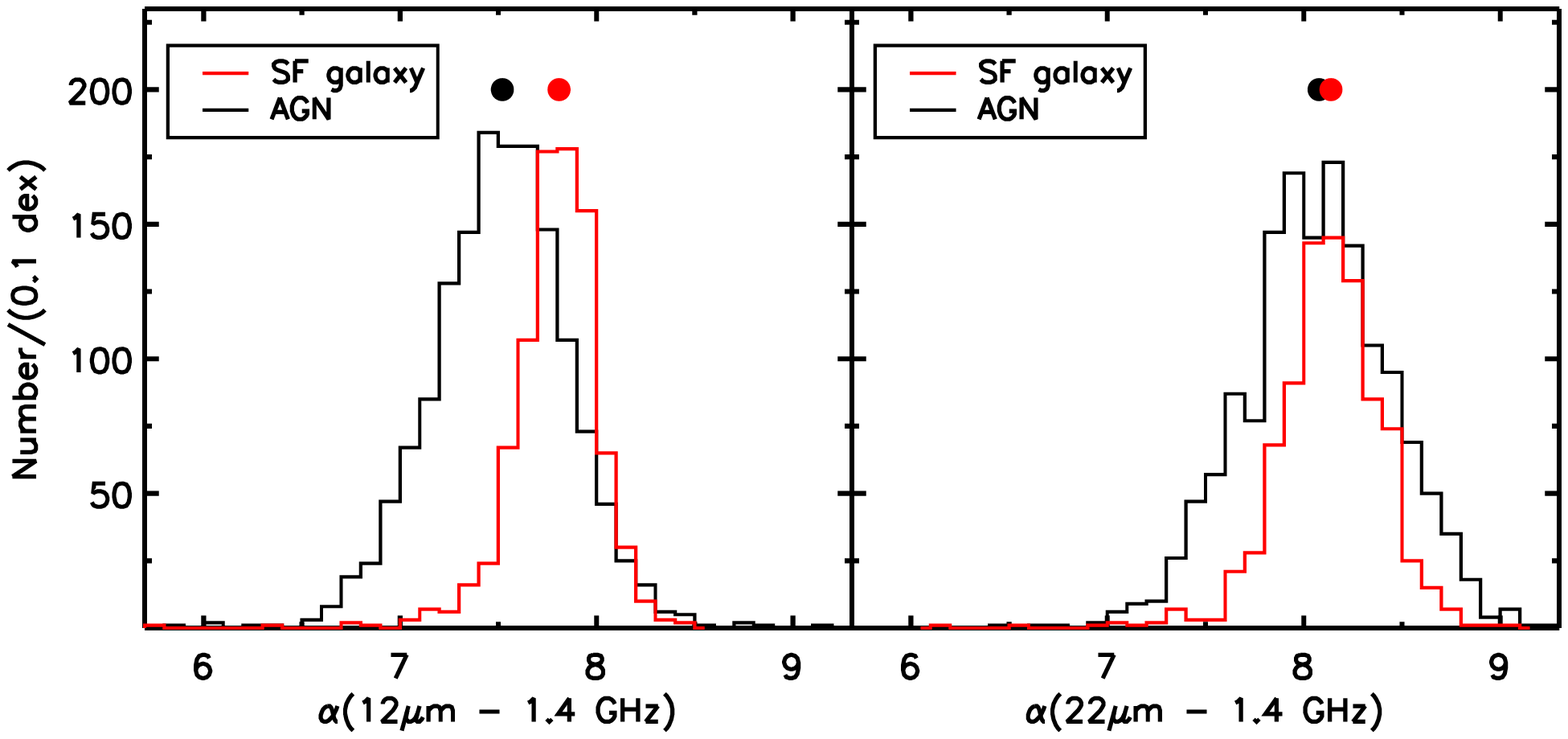}
\caption{Distributions of the rest-frame MIR-Radio spectral index $\alpha$, the ratio of the 12 \mics\ luminosity (left panel)
or 22 \mics\ luminosity (right panel) to the 1.4 GHz luminosity. The distributions of $\alpha$ for
AGNs on Branch A (black histograms) may be compared to those of mass-matched SF galaxies (red histograms).
The median values of each distribution are marked with appropriate red or black circular points at the top of each panel.
}
\end{figure*}

Here we investigate the likely
cause for the splitting of the AGN into two populations in MIR-Radio space. This has important
implications for our understanding of the AGN population. If the dichotomy is driven by differences in the level
of SF-heated dust, then our results would imply that the vast majority of Seyferts are in SF galaxies, even those with AGN-dominated
nuclear emission lines. If, on the other hand, the difference is driven by a higher level of AGN-heated
dust in Seyferts, then the MIR-Radio diagram may be viewed as a sensitive tracer of reprocessed high energy radiation from 
AGN and will allow us some insight into the intrinsic differences in the nuclear SEDs of these two classes of systems.
Our approach is to devise a series of empirical tests which make use of the high-quality ancillary data and measurements
available for SDSS galaxies.
 
\subsection{Star-Forming galaxies on the MIR-Radio plane}

A simple empirical diagnostic is the location of pure inactive, star-forming galaxies in MIR-Radio space.
AGNs are mostly found in fairly massive galaxies (\smass$\gtrsim 10^{10.5}$ \msun) and, since many
galaxy properties, including SFR, correlate strongly with stellar mass, a fair comparison of SF properties between AGNs
and inactive galaxies requires, minimally, that the comparison take into account stellar mass related biases 
\citep{silverman09, xue10, rosario13}. 
For this, we construct a control sample of pure star-forming galaxies matched to AGN in both redshift and \smass. We select
galaxies which lie below the curve that separates star-forming from composite systems
in the BPT diagram \citep[Eqn.~1 in][]{kewley06}. To ensure good discrimination in this diagram, we also require the
galaxies to be detected with a S/N$>3$ in the \ha, \hb\ and \ntwo\ emission lines. Only a valid upper limit is needed on \othree\ to
select SF galaxies in this manner. After binning the AGNs and inactive galaxies in $\Delta\log$\smass$= 0.1$ 
bins in stellar mass and $\Delta z= 0.01$ bins in redshift, we randomly assign one inactive galaxy to each AGN in a bin.
In practice, since there exist very few purely SF galaxies with large stellar masses, 
some AGNs are left unmatched at the high mass end (\smass$\gtrsim 10^{11.5}$ \msun).
We account for the unmatched AGNs in the following analysis. The inactive control sample was crossmatched to the
WISE All-sky survey catalog and the VLA-FIRST catalog in the same fashion as the AGNs, and the MIR and radio
photometry were extracted from these catalogs in a similar way.

In Figure \ref{lir_lradio_sfcomp}, we plot again the MIR-Radio diagrams for the AGN, as in Figure \ref{lmir_lradio_total}, but also include
the SF galaxies as red points. The SF galaxies describe a tight relation in MIR-Radio space. In the
\lmir-\lrad\ diagram (left panel), the SF galaxies overlap with the upper end of the distribution of AGNs on Branch A, but essentially no
inactive SF galaxies are found on Branch B. SF galaxies also lie exclusively on Branch A in the \lfir-\lrad\ diagram (right panel), but here
the SF galaxies generally overlap with the AGNs. In both panels, linear regression lines for AGNs and inactive
galaxies are shown, estimated using an ordinary-least-squares (OLS) bisector algorithm. Only the AGNs from Branch A that were
successfully matched to inactive galaxies are used in the regression analysis. 
The relative trends for both sets of objects can be compared at a glance.

In the right panel of Figure \ref{lir_lradio_sfcomp}, we include as well the ridge line of the FIR-radio correlation found among SF galaxies 
\citep[e.g.][]{condon92, sargent10}, extrapolated to rest-frame 22 \mics\ using the SF templates of \cite{ce01}. 
The increased contribution of warm dust in the MIR at the transition between normal SF galaxies and luminous IR galaxies (LIRGs),
as modelled by \cite{ce01}, is responsible for the kink in the shape of this line in this diagram; real galaxy populations
likely show a smoother relationship. Inactive SF galaxies and AGNs on Branch A scatter quite symmetrically 
about the FIR-radio correlation, implying that the 22 \mics\ luminosity of both is largely produced
by SF heated dust.

A closer look at the ratio of MIR-to-radio luminosity brings out the differences
at 12 \mics\ and 22 \mics. We define a MIR-Radio spectral index $\alpha = L_{\nu}\textrm{[MIR]}/L_{\nu}\textrm{[1.4 GHz]}$, where
the MIR$=$12 or 22 \mics. In Figure \ref{ir_radio_ratio}, we compare the $\alpha$ distributions of stellar mass-matched SF
inactive galaxies and AGNs on Branch A. The SF galaxies always show a narrower
distribution than the AGNs. While the median $\alpha$ values for both sets of objects are approximately the same at 22 \mics,
differing by 0.06 dex, the AGNs are weaker compared to the SF galaxies at 12 \mics\ by 0.29 dex. We discuss possible reasons
for these differences in Section 5. Since \lfir\ is more consistent with an origin in SF, we use only 22 \mics\ measurements in further tests.

\begin{figure}[t]
\figurenum{8}
\label{akari_sources}
\centering
\includegraphics[width=\columnwidth]{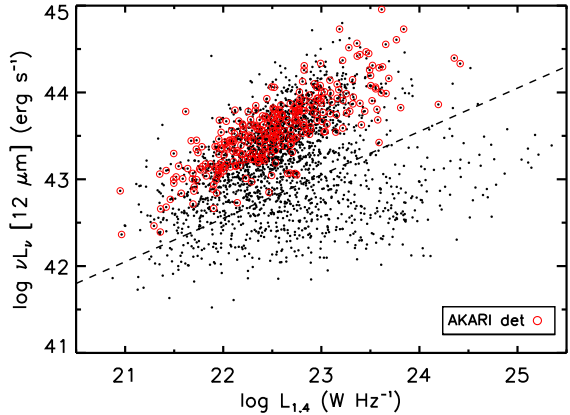}
\caption{Rest-frame 12 \mics\ luminosity (\lmir) vs. 1.4 GHz luminosity (\lrad) of SDSS selected AGNs (black points)
in the redshift range $0.02<z<0.15$. Sources detected at 90 \mics\ in the AKARI/FIS all-sky survey are marked
with open red circles. Every one of these FIR-bright SF host galaxies lies on Branch A.
}
\end{figure}

As a final test of the relationship between Branch A and star-formation, we plot the location of the AKARI 90\mics\ detected
sources on the \lmir-\lrad\ diagram for AGNs (Figure \ref{akari_sources}). Given the depth of the AKARI/FIS survey, the hosts of these
AGNs must be forming stars at a modest rate of several to tens of \msun/yr. 
Every single AKARI detected source lies on Branch A, which strongly reinforces the conclusion that this Branch marks the location
of SF host galaxies in the MIR-Radio plane.

\subsection{The relation between \dfour\ and MIR luminosity}

\begin{figure}[t]
\figurenum{9}
\label{d4000_lfir}
\centering
\includegraphics[width=\columnwidth]{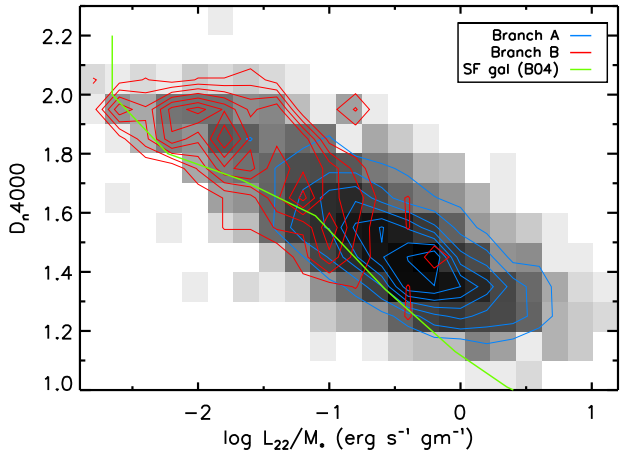}
\caption{ \dfour\ vs.~the ratio of \lfir\ to stellar mass (\smass) of  SDSS selected AGNs
in the redshift range $0.02<z<0.15$ plotted as a density map. A $\sinh^{-1}$ stretch is applied
to the map to enhance low density regions. A strong anticorrelation
is seen, as expected if \lfir\ is related to SFR. There is a marked difference in the location of objects 
from Branch A (blue contours) and Branch B (red contours) on this plot. The green solid line shows
the ridgeline (modal trend) of the correlation between \dfour\ and sSFR from \cite{brinchmann04},  
adapted to this Figure using a relation between SFR and \lfir\ derived from the calibration of
 \citet{kennicutt98} and the SED library of \citet{ce01}. 
}
\end{figure}

The optical spectral index \dfour, sensitive to the strength of the 4000\AA\ break, is a good measure
of the light-weighted specific SFR (sSFR) of galaxies \citep{brinchmann04}. If the two branches identified above do indeed correspond to 
AGNs with different levels of on-going SF in their hosts, one may expect a difference in the \dfour\ distributions of AGNs in the two branches. 
This is tested in Figure \ref{d4000_lfir}, where we plot \dfour\ against \lfir/\smass. If the 22 \mics\ luminosity is determined primarily
by the SFR of the AGN hosts, then \lfir/\smass\ will be proportional to the sSFR and, therefore, tracked by \dfour.
Indeed, we find a strong anticorrelation between \dfour\ and \lfir/\smass. In addition, objects from the two branches are
reasonably well separated in this diagram. Branch B AGNs have a higher \dfour, clustering around a value of 1.9
at which the sensitivity of \dfour\ as a tracer of stellar age saturates in old stellar populations. In contrast, objects in Branch A
show typically much smaller \dfour, peaking around a value of 1.4 and form much of the ridgeline of the correlation
in the Figure. Clearly, the separation of AGNs into two branches also largely separates them by sSFR. 

In addition to the AGNs, we also plot the modal trend between \dfour\ and sSFR from \citet{brinchmann04} as a green line
in the Figure. We have used the SED library of \citet{ce01}, along with the relation between SFR and $L_{FIR}$
from \citet{kennicutt98}, to convert between SFR and \lfir\ and place the  \citet{brinchmann04} line on this Figure.
At a given \lfir/\smass, AGNs have a higher \dfour\ than those of SF galaxies, irrespective of the Branch to which
they belong. We have checked that galaxies from our SF inactive control sample lie along the \citet{brinchmann04} line, verifying
that the offset we see for AGNs in not due to systematic differences in the estimation of the SFR from \lfir\ (\citet{brinchmann04}
calibrate SFRs from \ha). Since we have shown that the \lfir\ of AGNs and SF galaxies are not systematically different, certainly
not at the level of $\approx 0.3$ dex needed to reconcile the AGNs and SF trends in Figure \ref{d4000_lfir}, this would
suggest that AGNs have higher \dfour\ than SF galaxies of similar stellar mass, or, alternatively, AGNs show a lower sSFR
than SF galaxies. At face value, this is puzzling, given the similarity in the galaxy integrated SFRs of the two populations
as traced by their 22 \mics\ luminosity distributions. However, the difference makes more sense if we recall the aperture-related 
selection effects of SDSS spectroscopic samples.  BPT-selection of AGN-dominated systems systematically selects \emph{against} 
objects with strong SF in the SDSS fiber aperture, while the selection of pure SF galaxies
selects \emph{for} strong central SF. 
Against the backdrop of the older bulge
stellar populations found in massive galaxies, this selection effect will naturally lead to lower measured sSFRs for the AGNs.  
 
\subsection{Radio-loudness and its relation to the Branches}

\begin{figure}[t]
\figurenum{10}
\label{d4000_lradio}
\centering
\includegraphics[width=\columnwidth]{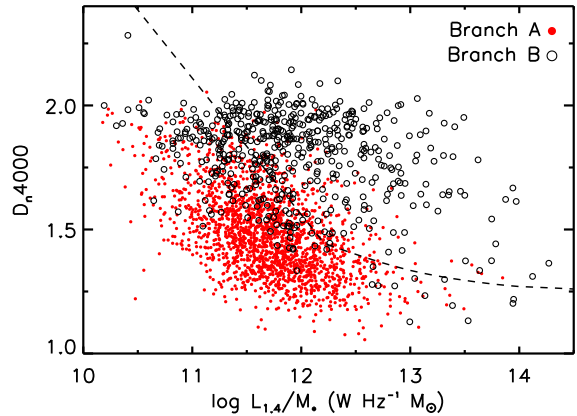}
\caption{ \dfour\ vs.~the ratio of \lrad\ to stellar mass (\smass) of  SDSS selected AGNs
in the redshift range $0.02<z<0.15$. The dashed line is from \citet{best12}; radio-loud sources lie above
the line in this diagram. There is a marked difference in the location of objects 
from Branch A (red filled points) and Branch B (black open circles). The radio-loud region of the
diagram is occupied mostly by objects on Branch B.
}
\end{figure}

Further examination of Figure \ref{lmir_lradio_total} reveals a interesting connection between the strength
of radio emission and the division into branches. Consider the small number of sources with \lrad$>10^{24}$ \whz, 
which are too radio-loud to be powered by even the most luminous SF galaxies. 
We find that most of these sources lie on Branch B, with relatively few radio galaxies lying on Branch A. We can
explore this connection more robustly by plotting \dfour\ against the specific radio luminosity (\lrad/\smass)
of our AGNs (Figure \ref{d4000_lradio}). This plot was first proposed in \citet{best05a} and recalibrated
in \citet{best12} as a way to identify sources with a strong radio excess over that produced by SF. Pure star-forming 
galaxies with a range of SF histories lie along a fixed locus in this diagram and below the dashed line in Figure \ref{d4000_lradio}. 
Above the line, sources may be treated as radio-loud. 

As one may see at a glance, sources on Branch A lie primarily in the radio-quiet, SF-dominated part of the diagram, with
only 12\% lying above the dashed line. Sources on Branch B straddle the line, but most are in the radio-loud part of the diagram.
As hinted by Figure \ref{lmir_lradio_total}, radio-loud AGNs in our sample lie mostly on Branch B. The scatter
about the separation line depends on a complex set of factors such as the biases of BPT AGN selection (Section 2.1.1)
and SDSS aperture effects, since \lrad\ tracks the galaxy integrated radio luminosity, while \dfour\ only traces
the inner stellar populations. However, we take the broad separation of the branches in this diagram as evidence that
radio-loud AGN lie preferentially on Branch B, while Branch A is preferentially populated by radio-quiet SF hosts.

The dearth of SF signatures and MIR weakness in powerful radio galaxies (with \lrad$>10^{26}$ \whz)
is well-known from previous studies \citep[e.g.][]{ogle06, shi07, dicken12}. 
Comparisons of radio-loud quasars and radio galaxies also indicate high MIR
optical depths in the latter \citep{leipski10} and boosted non-thermal emission in the former \citep{cleary07}, both of
which must be considered in a complete picture of the relative MIR-to-radio properties of radio-loud systems.

\subsection{AGN-heated dust in the MIR}

\begin{figure*}[t]
\figurenum{11}
\label{lo3_lfir}
\centering
\includegraphics[width=\textwidth]{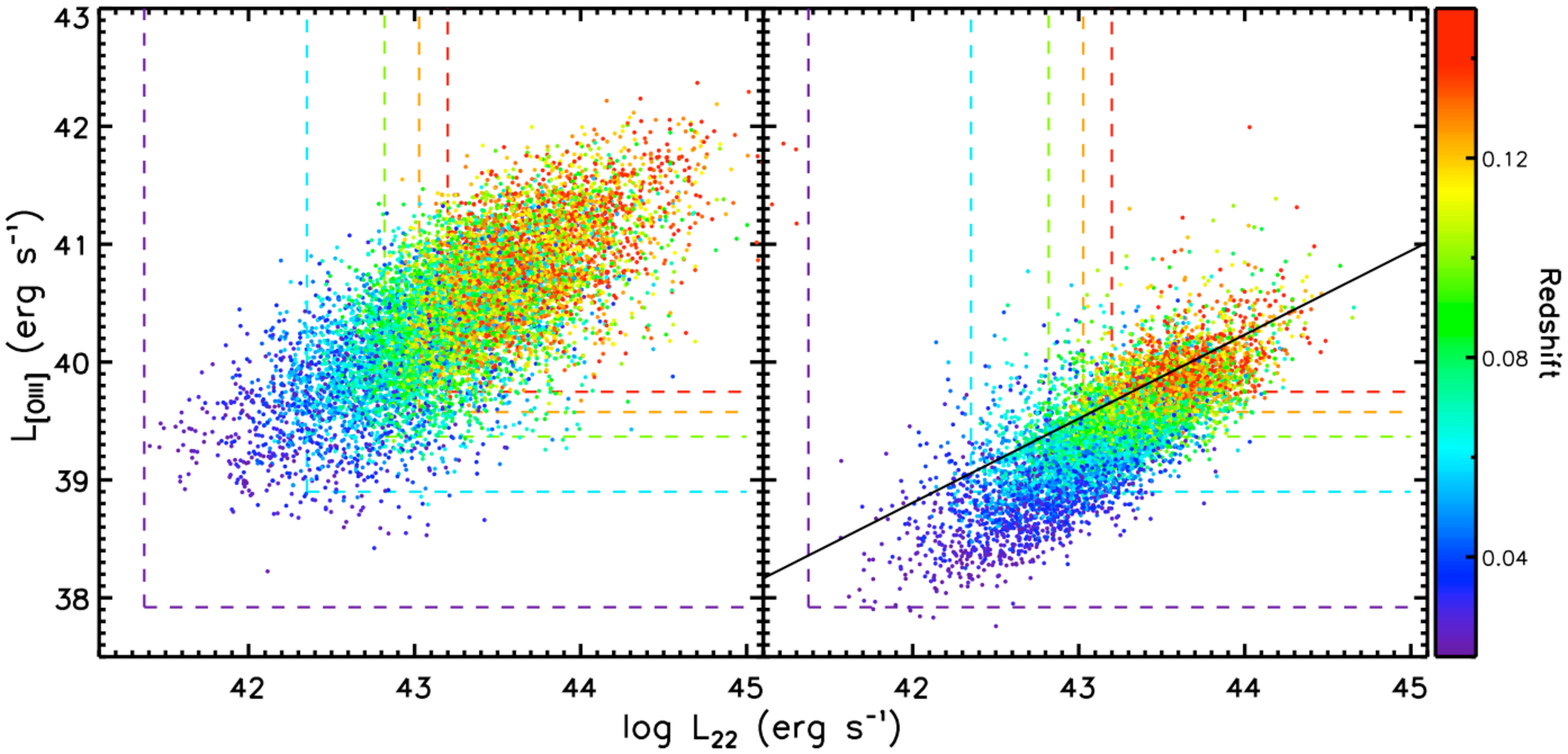}
\caption[]
{\othree\ luminosity (\lothree) vs.~rest-frame 22 \mics\ luminosity (\lfir) of SDSS emission-line selected AGNs (left panel)
and pure SF galaxies (right panel) colored by redshift, according to the colorbar at right. 
The dashed lines show the typical luminosity limit of the WISE and SDSS spectrosopic surveys 
at a set of illustrative redshifts, also colored according to the same colorbar.
Both sets of galaxies show a correlation between \lothree\ and \lfir. 
For a given \lfir, SF galaxies have much lower \lothree\ and tend to cluster close to the SDSS 
luminosity limits at all redshifts. The solid black line in the right panel is an envelope that contains below 
it 80\% of all SF galaxies in this diagram.

}
\end{figure*}

\begin{figure}[t]
\figurenum{12}
\label{visir_agn}
\centering
\includegraphics[width=\columnwidth]{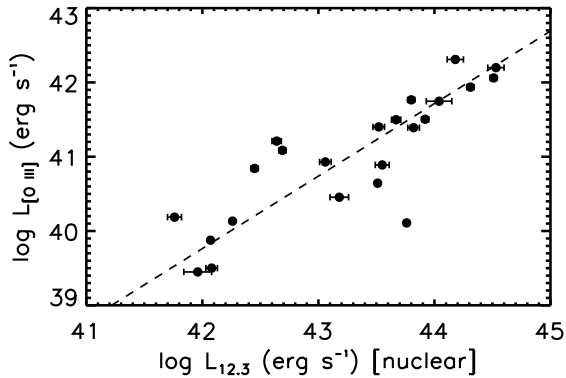}
\caption[]
{The correlation between the nuclear 12.3 \mics\ luminosity (measured within 1") and the integrated \othree\ luminosity
from the Narrow-Line Region of a set of nearby AGNs. The dashed line is a OLS bisector fit to the data points. See
Section 4.3 for details.
}
\end{figure}

\begin{figure*}[t]
\figurenum{13}
\label{mixing_diag}
\centering
\includegraphics[width=\textwidth]{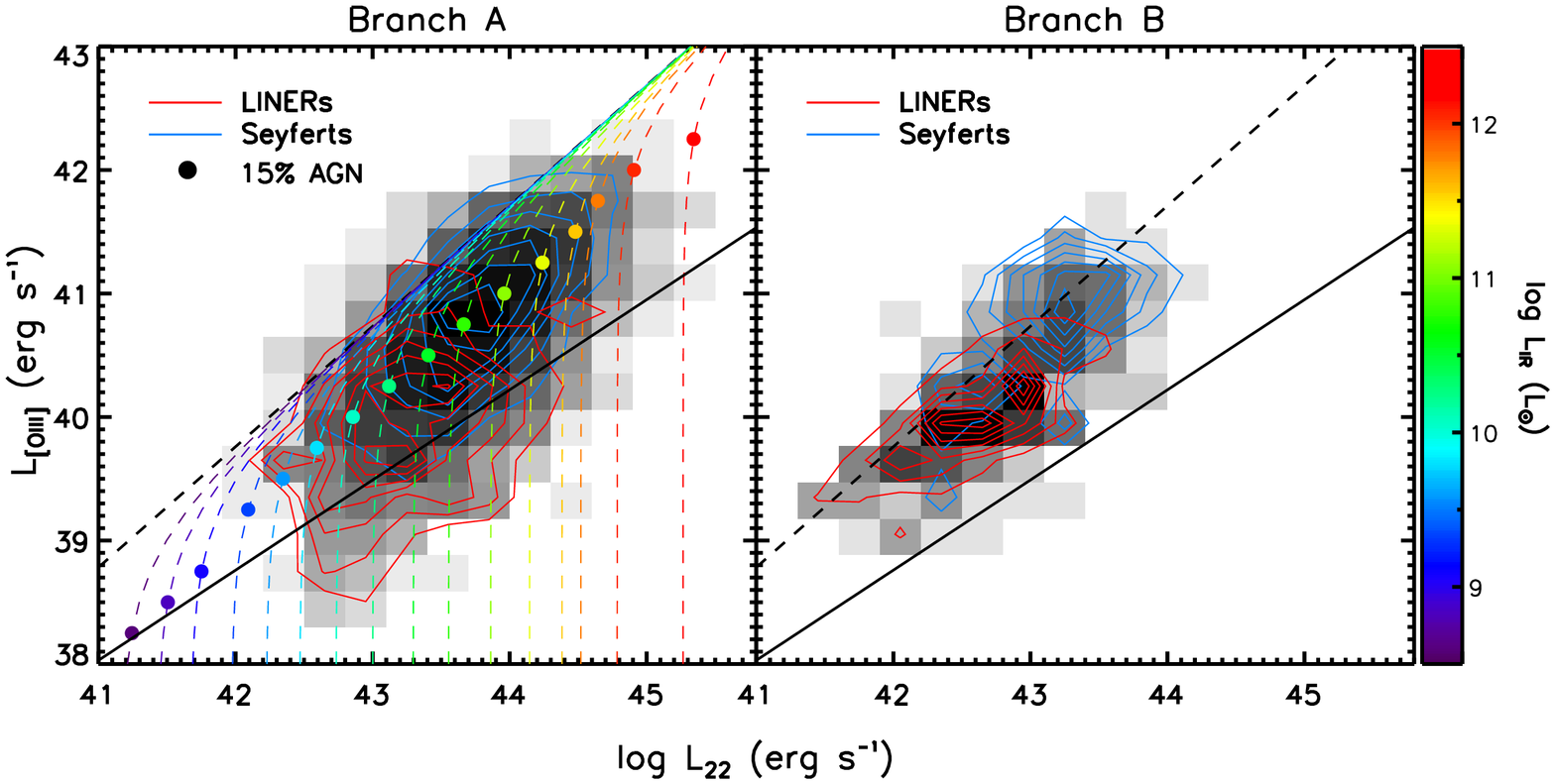}
\caption[]
{\othree\ luminosity (\lothree) vs.~rest-frame 22 \mics\ luminosity (\lfir) of SDSS emission-line selected AGNs 
on Branch A (left panel) and Branch B (right panel), plotted as a density map. A $\sinh^{-1}$ stretch is applied
to the map to enhance low density regions. In each panel, the locations of
Seyferts (blue contours) and LINERs (red contours) are also shown. The solid black line is the 80\%
envelope of SF galaxies (see Figure \ref{lo3_lfir}), while the dashed black line shows the location
of a pure AGN template on this diagram. The colored dashed lines in the left panel show the loci
of hybrid templates of star-forming galaxies containing AGN, with a SF-powered IR luminosity
indicated by the colorbar at right. As the AGN luminosity traced by \lothree\ reaches
a level where it starts to dominate the MIR luminosity of a hybrid template, the loci transition from a fixed \lfir\
to a sloped line on which AGN-dominated systems lie. The colored filled circles mark the location on the
hybrid template loci where the AGN accounts for 15\% of \lfir. Since the MIR in AGNs on Branch B is not governed
by star-formation, hybrid templates are not plotted in the right panel. 
}
\end{figure*}

Till now, our tests have connected star-formation and the emission of AGN hosts in the long-wavelength
MIR. However, given the expected prominence of AGN-heated dust at these wavelengths, 
it is important as well to constrain the contribution from the AGN. 

\subsubsection{The \lfir-\lothree\ relation in AGNs and star-forming galaxies}

A simple test for the influence of the AGN in the MIR is shown in the left panels of Figure \ref{lo3_lfir}, where we plot \lfir\ against \lothree\
for all WISE-detected AGNs, including those that are not detected in the FIRST survey.
In AGN-dominated galaxies, the [O III] emission lines are a good measure of the intrinsic luminosity of the AGN, since they originate
in the highly ionized Narrow Line Region (NLR). In this and following plots, we only use the observed \othree\
luminosities without applying any correction for NLR dust extinction. This is because the accuracy of an extinction correction
from the \ha/\hb\ Balmer decrement, is highly dependent on the flux uncertainties of the \ha\ and \hb\ 
emission lines.
This will lead to inaccurate corrections for objects with fainter emission line fluxes, such as low luminosity and distant AGNs. 
Therefore, and also for consistency with the compilation of [O III] luminosities discussed below, we present
our analyses using uncorrected \lothree. However, we have tested our results using a fixed extinction of $A_{V}=0.8$
the typical value estimated for the brighter AGNs in our sample. The basic conclusions remain unchanged.

We find that AGNs show a good correlation between \lothree\ and \lfir. The correlation suggests either that the AGN
contributes significantly to the luminosity of host galaxies at 22 \mics, or, alternatively, that the total
SF in AGN-dominated galaxies tracks nuclear activity. Similar results have been shown before
for AGN-dominated galaxies from the SDSS using other tracers of the SFR, such as \dfour\ \citep{netzer09}.
We discuss these alternatives in more detail in Section 5.3.

Note that purely SF galaxies from our control sample also show a strong correlation between \lothree\
and \lfir, as demonstrated in the right panel of Figure \ref{lo3_lfir}. The slope of this correlation is similar to that shown by
AGNs, but it is shifted towards lower \lothree\ by 1.7 dex. Given the characteristic flux limits of the SDSS spectra, which
are translated to luminosity limits at different redshifts and plotted as dashed lines in the Figure, one concludes that, 
the correlation seen in SF galaxies is mostly shaped by Malmquist bias. At progressively higher redshifts,
only galaxies luminous at both 22 \mics\ and in [O III] will be selected, tightening the correlation.
Nevertheless, we can use our sample to constrain the degree to which emission from SF may contribute
to the \othree\ emission in the AGNs. We construct an envelope which contains 80\%\ of SF galaxies, shown
as a solid line in the Figure, which we compare below to the location of the AGNs on this diagram.

\subsubsection{The intrinsic MIR luminosity of AGNs}

Before we can embark on this exercise, we require a predictor for the intrinsic MIR luminosity of the AGN which we can compare
to the measured MIR luminosity of the WISE-detected sources. Many studies have used MIR spectroscopy to empirically
investigate the relation between nuclear power and the MIR continuum or PAH luminosity in nearby and distant AGNs
\cite[e.g.][]{lutz04,shi07,diamond-stanic10,lamassa10}. Most studies rely on fairly large apertures in the MIR -- for e.g., with the Spitzer
IRS spectrograph, the workhorse for most of such studies, typical apertures used to extract spectra correspond to scales of a few
to several kpc. The contribution of SF-heated dust to the long-wavelength MIR in such spectra could be substantial. 
Therefore, we turn to a 
high resolution (sub-arcsec) photometric study of the MIR emission in local AGNs using narrow-band imaging from 
the VISIR instrument on the VLT \citep{gandhi09, asmus11}. These small-scale observations greatly limit host galaxy
dilution and isolate, as best as currently possible, the AGN-heated torus emission from the nucleus. Having said this,
in some cases the VISIR measurements may still have a substantial SF contribution from starbursts on tens of pc scales \citep{asmus11}.

For a set of \othree\ emission line fluxes, we employ the compilation of \citet{whittle92}, which uses a coherent 
methodology to account for aperture losses, bringing consistency to the varied nature of spectroscopic 
studies of AGNs in the Local Universe. \citet{whittle92} only provide observed line fluxes, uncorrected
for extinction, but since we use uncorrected \lothree\ in our full analysis, we directly adopt these measurements. 
Combining 12.3 \mics\ luminosities with \lothree\ for Seyferts in common to 
the two datasets, we establish a tight correlation between these quantities (Figure \ref{visir_agn}) which we fit using an 
OLS bisector regression algorithm. The correlation is tighter for the more luminous AGNs, but the scatter 
increases at lower luminosities, possibly due to a higher level of SF contamination among such systems. 
Nevertheless, we use our best-fit relationship to estimate a 12.3 \mics\ luminosity for an AGN given \lothree:

\begin{equation}
L_{12.3} \; = \; 1.02 L_{[OIII]} + 1.24
\end{equation}

Armed with this relationship, we construct a set of hybrid IR templates of AGNs embedded in star-forming hosts. 
The average MIR template of Type I AGNs from \citet{mor12} is combined with the SF SED library of \citet{ce01}, scaling
the AGN template using Equation 1 to cover a range of nominal AGN luminosities in the range 
\lothree$=10^{38\textrm{--}43}$ \ergs. The resulting set of hybrid templates span a wide range in MIR
AGN dominance, from sub-percentile to 100\%\ AGN emission at 22 \mics. There is considerable
scatter of both real galaxy and AGN SEDs about our adopted templates, so these models are not exact for
any individual object, but will serve as a guide to the typical behavior of star-forming AGN hosts with varying 
degrees of nuclear emission.

\subsubsection{The star-formation/AGN mixing diagram}

In Figure \ref{mixing_diag}, we again plot \lothree\ against \lfir, now only for sources with both FIRST and WISE detections, i.e.,
the subsample which can be classified into branches. 
The distribution of AGNs are shown as a density map, but now splitting the AGNs by panel into the two branches. 
Overplotted on the left panel over the distribution for Branch A
is the loci of the hybrid templates, shown as dashed lines colored by SF-powered IR luminosity. These
lines are not plotted on the right panel since 
the MIR luminosity of the objects in Branch B probably does not arise from recent star-formation.

The hybrid template loci have a characteristic shape: as the AGN luminosity in a particular hybrid increases, 
it moves along a track of increasing \lothree\ at a fixed \lfir\ till the AGN fraction at 22 \mics\ starts to approach 
unity. At this point, all loci bend onto the pure AGN line (dashed line in both panels). Naturally, the characteristic AGN \lothree\ at the 
turnover is a function of the IR luminosity of the SF template. Interestingly, among the luminous IR galaxies 
(with $L_{IR} > 10^{11}$ \ergs), it is possible to conceal the MIR emission of quite a luminous AGN, 
with \lothree\ as high as $10^{41}$ \ergs\ \citep[$L_{bol} \approx 4\times10^{44}$ \ergs, based on the
bolometric correction of][]{heckman04}. 

A key point to take from Figure \ref{mixing_diag} is that the \lfir\ of pure AGN templates is quite close to the observed \lfir\ for
the AGNs, given their [O III] luminosities. 
Disentangling SF and nuclear components of the MIR luminosity in these AGNs is therefore not trivial. Nevertheless, we proceed
given our assumptions about the intrinsic AGN luminosity and the SEDs of AGNs and SF components, and discuss
some of the complexities in Section 5.

AGNs from both branches show a correlation between \lfir\ and \lothree. In detail, there are some differences between
the behavior of the two branches in the diagram. The characteristic \lothree\ of AGNs in Branch B is lower than that for 
Branch A, a consequence primarily of the larger fraction of low-luminosity LINERs in Branch B. 
The ridgeline of the Branch B distribution lies closer to the pure AGN locus than that of Branch A,
as can be judged from the 
sequence of large colored circular points plotted in the left panel of the diagram.
The points show the location on the template loci where the AGN accounts for 15\% of \lfir, 
chosen to run through the peak of the Branch A distribution. Clearly the sources on Branch B are
more AGN dominated in the MIR, even though they contain less luminous AGNs. 
Having said this, we see that the ridgeline of the Branch B distribution is still slightly offset from the pure AGN locus, most
clearly for the LINERs. This could reflect the presence of additional sources of weak MIR emission
such as cirrus heating by evolved stars \citep{sauvage92,calzetti95,groves12} or the synchrotron tail 
from relativistic particles \citep[e.g.][]{yuan07} which only become prominent in low-luminosity systems with minimal SF and AGN heating.

In both panels of Figure \ref{mixing_diag}, we use different colored contours to show the location of Seyferts and 
LINERs in each panel. As expected, Seyferts have higher \lothree\ than LINERs in both branches. In fact, some LINERs
on Branch A are weak enough in [O III] to place them in the region occupied by the star-forming galaxies. A substantial
fraction of the line emission in such systems could arise from HII regions. In addition to this, the
LINERs on Branch A show a weaker correlation between \lothree\ and \lfir\ than the Seyferts on the same Branch.
Their typical location on the diagram is also further away from the AGN line than the Seyferts, indicating
that such star-forming LINERs are even more dominated at 22 \mics\ by SF-heated dust. 

It is worth noting that the median \lothree\ of LINERs on Branch A and Branch B are similar
despite an order of magnitude difference in their median \lfir. This also highlights the conclusion that the separation of
the AGN population into branches is mostly governed by processes unrelated to their direct nuclear emission.

\section{Discussion}

\begin{figure}[t]
\figurenum{14}
\label{scenarios}
\centering
\includegraphics[width=\columnwidth]{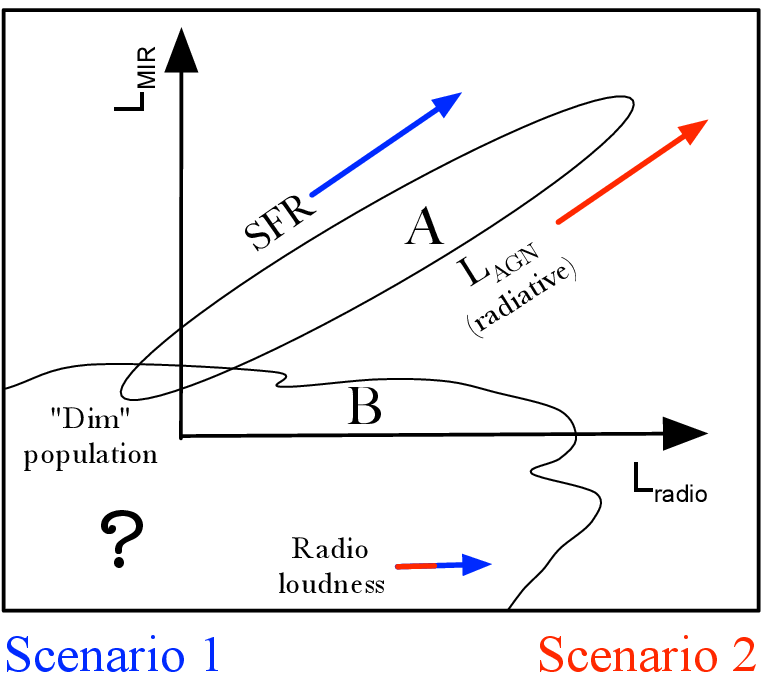}
\caption[]
{ Schematic descriptions of the two scenarios for the origin of the branches in the MIR-radio plane outlined in Section 5. 
The two branches are labelled `A' and `B' as from Figure \ref{lmir_lradio_total}. The ``dim" population which
lie below the joint WISE and FIRST detection limits, is designated with a question mark signifying our lack of knowledge of their
true MIR-radio relationship. Scenario 1, denoted in blue, assumes that star-formation purely governs the location of AGN hosts on Branch A. Scenario 2, denoted in red, assumes that pure AGN emission governs all patterns on the MIR-radio plane. See Section 5.2 for more details and a discussion.

}
\end{figure}

\subsection{Summary of empirical results}

The patterns exhibited by AGNs in Figures \ref{lmir_lradio_total} and \ref{lmir_lradio_separate} strongly suggest the existence
of distinct populations (branches) of active galaxies, differentiated by their MIR-to-radio properties. Intimately connected to the nature
of the branches is the origin of the integrated (i.e., galaxy-wide) MIR emission of the AGNs.
Before proceeding with a interpretive discussion, we summarize what we have learned about the branches from the various
tests performed above. 

Sources on {\bf Branch A}: have higher absolute MIR luminosities than those on Branch B (Figure \ref{lmir_lradio_total});
describe a steep trend in MIR-radio space, shared with SF galaxies (Figure \ref{lir_lradio_sfcomp});
account for 95\% of Seyfert AGNs and 50\% of LINERs (Figure \ref{lmir_lradio_separate});
are consistent with the median MIR-radio relationship of SF galaxies at 22 \mics\ (Figure \ref{ir_radio_ratio});
are offset low from the median MIR-radio relationship of SF galaxies at 12 \mics\ by $\approx 0.3$ dex (Figure \ref{ir_radio_ratio});
have a larger scatter in their MIR-radio distribution compared to SF galaxies (Figure \ref{ir_radio_ratio});
include all FIR-bright AGN hosts with substantial SF (Figure \ref{akari_sources});
have a low median \dfour\ $\approx 1.4$ (Figure \ref{d4000_lfir}); 
 are largely radio-quiet, with radio luminosities consistent with a SF origin (Figure \ref{d4000_lradio}); 
 are consistent with typically low AGN contributions at 22 \mics\ of 15\%\ (Figure \ref{mixing_diag}).

In comparison, sources on {\bf Branch B}: have lower absolute MIR luminosities than those on Branch A (Figure \ref{lmir_lradio_total});
describe a shallow trend in MIR-radio space, consistent with uncorrelated scatter modulated by redshift-dependent 
MIR luminosity limits (Figure \ref{lmir_withz});
account for 50\% of LINERs and only 5\% of Seyferts (Figure \ref{lmir_lradio_separate});
lie well off the MIR-radio relationship of SF galaxies (Figure \ref{lir_lradio_sfcomp});
have a higher \dfour\ $=[1.6,2.0]$ (Figure \ref{d4000_lfir});
account for most radio-loud AGNs (Figure \ref{d4000_lradio});
are more AGN-dominated in the MIR than Branch A (Figure \ref{mixing_diag}).

\subsection{Origin of the Branches in the MIR-Radio plane}
  
Our investigation has revealed evidence for a mixture of different heating mechanisms for dust in the hosts of AGNs: 
star-formation, nuclear light and possibly a contribution from cirrus heated by the UV background in galaxies. 
Here we critically examine the role of these mechanisms in the nature of the branches. First we consider 
two simplified alternate explanations, which allow us to conceptually explore the assumptions and conclusions
one may draw from the empirical analyses of the last two sections. Then we consider the consequences of
combining these alternatives towards a more realistic and nuanced picture of the MIR properties of AGNs.

We begin by outlining two extreme and opposite positions: 1)  The 12--22 \mics\ luminosity of AGNs on Branch A is dominated by 
SF-heated dust, or, 2) The 12--22 \mics\ luminosity of AGNs on both branches are dominated by
AGN-heated dust. These scenarios are outlined schematically on the MIR-radio plane in Figure \ref{scenarios}. 
In both cases, we assume there is a substantial population of non star-forming, low-luminosity
AGN hosts most of which lie below the joint SDSS, WISE and FIRST detection limits. We term this the ``dim"
population, signifying their weakness in both SF and nuclear emission. We do not speculate 
here on the MIR-radio relationship of the dim population, representing them instead 
as a cloud with a question mark signifying our lack of knowledge. Only a small fraction, the tip
of the iceberg, are detected in our sample, perhaps due to higher than average levels of dust in their
hosts heated by the AGN or the diffuse UV field, or possibly shocks from radio jets. 
These constitute the systems on Branch B. Fully detected AGNs lie within the quadrant delineated 
by the axes in Figure \ref{scenarios}. 

\subsubsection{Star-Formation as the origin of the Branches}

This scenario is represented using blue arrows and lines in Figure \ref{scenarios}. 
In addition to the dim population described above, there is a set of AGNs found in galaxies with wide-spread star-formation. 
These galaxies will have elevated IR and radio luminosities over the dim population 
and lie along the correlation between IR and radio continuum luminosities. They constitute Branch A.

Support for this scenario comes from the various relationships between the nature of the branches and measures
of SF. Objects on Branch A show a tight correlation between \lfir\ and \dfour, while most Branch B AGNs lie in evolved systems
(Section 4.2). FIR-bright AGNs all lie on Branch A (Section 4.1). But the most crucial constraint comes from the 
correspondence between the \lfir--\lrad\ relationship of Branch A and that of pure SF galaxies. This correlation is widely 
understood to be set by the close relationship between the non-thermal output from supernovae and the UV 
emission from young stars, reprocessed to the IR. If AGN-heated dust significantly affected \lfir, we would have seen
an offset towards higher 22 \mics\ luminosities among the AGNs, which we do not observe, except perhaps among
some of the brightest AGNs in our sample. This, as well as the constraints from hybrid templates and the radio-loudness test, 
suggests that the MIR luminosity, as well as the radio luminosity, of most AGNs on Branch A is dominated by SF.

Interestingly, the 12 \mics\ luminosity of AGNs on Branch A is too low at a given \lrad\ to be fully consistent
with SF galaxies. In other words, AGNs on Branch A have redder W3-W4 colors than SF galaxies of the same 
(radio-based) SFR. Two effects may account for this. One is the suppression of PAH excitation or the
destruction of PAH grains in AGN-dominated systems, as suggested by recent
Spitzer/IRS spectroscopic studies of large samples of SDSS AGNs \citep{lamassa12}. Additionally, 
strong 10 \mics\ Si absorption troughs, frequent in Type II AGNs \citep{shi06, hao07}, could contribute to the offset.


\subsubsection{AGN emission as the origin of the Branches}

In Figure \ref{scenarios}, red arrows and lines represent the scenario where the hot-dust emission from the AGN torus
ultimately sets the patterns in the MIR-Radio plane. 
For this to be the case, our estimate of the intrinsic MIR AGN luminosity would have to
be systematically in error, to account for the $\sim 0.7$ dex in \lfir\ needed to reconcile the ridgeline of the
AGN trend with the AGN-dominated line in Figure \ref{mixing_diag}. 
This estimate is based on high resolution 12 \mics\ photometry of very nearby AGNs
combined with the best current knowledge of their intrinsic MIR SEDs.
It may be that the low to moderate luminosity AGNs in our sample have a redder mean MIR SED than 
existing studies currently suggest, leading to a more luminous 22 \mics\ luminosity than we apply to our hybrid
templates. This is supported by studies of the typical SEDs of low luminosity AGNs \citep[e.g.][]{ho08}, though
at the higher AGN luminosities among our sample, the \citet{mor12} SED should be accurate.

In this scenario, the two branches represent a bimodal distribution of AGNs with different levels of radio output
relative to their thermal (accretion disk) output. This is akin to the well-known ``Radio-loud/Radio-quiet" dichotomy 
\citep[e.g.][]{sikora07}. Objects on Branch B are low luminosity AGNs that span a range in radio luminosity 
and may be fueled by so-called ``hot-mode" accretion, while objects on Branch A, which contain most 
luminous AGNs and Seyferts, have a more definite relationship between their thermal and non-thermal 
output and are fueled mostly by ``cold-mode" accretion \citep{best12}.

An AGN-dominated MIR SED may help explain the redder W3-W4 color of AGN hosts compared to SF galaxies, since
pure AGN SED templates, such as that of \citet{mor12}, which lack PAH emission bands, are usually redder than the 
templates of weakly SF galaxies. Having said this, SF-related features, such as low ionization emission lines and PAH 
bands are known to be fairly common in the MIR spectra of AGN hosts \citep[e.g.][]{weedman05, buchanan06, wu09}, so pure 
AGN-dominated MIR SEDs in all systems does not have much empirical support. 
Earlier studies have reported a higher average AGN contribution to the MIR than we find \citep[e.g., $>45$\% among Type II AGN
at 19 \mics\ in][]{tommasin10}, but this is almost certainly because most existing spectroscopic
samples target rather nearby systems and sample only the inner few kpc of the host galaxy (Figure \ref{iras_comp}), 
missing most of the continuum and PAH luminosity emitted by SF on larger galactic scales. 

Besides this, an AGN-dominated scenario would still have to explain the close relationships found between
the branches and SF indicators, as well as the conspiracy that the \lfir--\lrad\ of the AGNs on Branch
A, set only by the intrinsic AGN output in this picture, overlaps that of pure SF galaxies. The processes that govern
the relative non-thermal output in AGNs (e.g., the Poynting flux from the magnetosphere of an SMBH)
is drastically different from those that set the non-thermal output of SF regions (i.e., shock acceleration in supernova
remnants), so a common slope and normalization for the trends in the MIR-Radio plane is very unlikely.

\subsubsection{A mixed origin}

In our analysis of the \lothree--\lfir\ diagram, we show that both SF and AGN-heated emission are expected to
produce luminosities close to the values seen among real AGNs. This suggests that the MIR luminosity of
AGNs, especially those on Branch A, is likely a mixture of AGN-heated and SF-heated components. The
investigation using hybrid templates suggests that the component from SF probably dominates by a small margin,
accounting for, on average, $\sim85$\% of \lfir. However, the AGN-heated component sets a floor to the
MIR luminosity -- even if the SFR of an AGN host galaxy is low, its total MIR luminosity cannot drop below the level set by
the nuclear component.

A mixed origin for Branch A could contribute to the broader width
of the $\alpha$(22 \mics$-$1.4 GHz) distribution for AGNs on this Branch (Figure \ref{ir_radio_ratio}). The SFR of the
host galaxy places a particular AGN at a particular location on the locus of SF galaxies in the MIR-Radio plane.
AGN activity can potentially boost both the MIR and radio luminosity over the level set by the SFR. If the AGN is radio-weak
but IR bright, it will scatter the system above the Branch A ridgeline. Alternatively, if the AGN is relatively radio-bright, it
will scatter the system below the ridgeline by increasing its \lrad\ without changing its MIR luminosity significantly. Differing
combinations of radio and MIR luminosity will scatter objects off the SF locus, leading to the broader distribution we see
among AGNs. The slight asymmetry in the shape of the $\alpha$(22 \mics$-$1.4 GHz) distribution may indicate that the
scatter from enhanced radio emission is larger than the scatter from AGN MIR emission. However, further careful study
are necessary to test this notion. 

\subsection{The Star-Forming Properties of AGNs}

One of the most interesting results from our study is that 95\% of SDSS/WISE/FIRST Seyfert galaxies 
lie on Branch A, while lower luminosity LINERs are more frequently on Branch B. We have shown that most objects
on Branch A have considerable on-going SF. This implies a relationship between the ionization
of AGN emission lines and star formation in their hosts. 

The characteristic low \lothree\ of LINERs imply that they are low-luminosity AGNs. Studies
have shown that they are typically in massive, evolved hosts with low levels of SF \citetext{e.g. \citealp{ho08}; but see \citealp{tommasin12}}.
An examination of Panel 1 in Figure \ref{lmir_lradio_separate} suggests that a fraction of LINERs have on-going
SF, though their typical SFRs, as tracked by their MIR luminosity on Branch A, are lower than in Seyferts. 
Combined with the fact that a large fraction of LINERs lie on Branch B, our results add support to the
view that LINERs are found mostly in quiescent or quenching hosts.

On the other hand, the very high fraction of Seyferts on Branch A is evidence that most Seyferts
are in actively star-forming host galaxies. This is consistent with GALEX-based UV studies that show that so-called ``bright AGNs"
in the SDSS are found among fairly normal SF galaxies \citep{salim07}. Since the ionization state of AGNs is
strongly correlated with \lothree, such bright emission-line selected AGNs are essentially all Seyferts. Bright
AGNs are known to have lower \dfour\ than lower-luminosity AGNs, indicative of younger mean stellar ages 
\citep{kauffmann03}. Morphological studies also reveal that Seyfert hosts are generally in massive galaxies with 
disks and active SF, while early-type galaxies hosting bright AGNs frequently show anomalously blue colors suggestive 
of on-going SF \citep[e.g.][]{schawinski10}. Studies of X-ray selected AGNs from deep extragalactic survey fields
also show a high frequency of SF host galaxies, compared to similarly massive inactive galaxies \citep{silverman09, rosario13}.
This suggests that radiatively efficient nuclear activity, which powers X-ray and emission-line bright AGNs,
is associated with on-going SF in the host galaxy. This can be explained by many diverse models, 
such as connected nuclear starbursts and AGN fueling \citep{davies07}, 
merger-induced AGN activity \citep{hernquist89}, accretion from stellar mass loss \citep{ciotti07}, or their
combinations. Alternatively, since both efficient nuclear fueling and SF rely on the same supply of material,
namely cold gas, the connection seen here may be simply governed by the stochastic availability of this fuel, rather than
any real causal link between the AGN and galaxy-wide SF \citep[e.g.][]{rosario13}. 

The correlation between \lothree\ and \lfir\ for objects on Branch A (left panel of Figure \ref{mixing_diag}) also seems
to suggest a correlation between the global SFR of AGNs and their nuclear luminosity. 
However, as described in Section 2.1.1, an important selection effect inherent to AGN-dominated systems 
selected on the BPT diagram, can play a role in strengthening this correlation, if not driving it completely.
Thus, we may expect to see a relationship between \lothree\ and \lfir\ even if a strong
correlation does not physically exist. Obviously a more complete assessment of correlations 
between the SFR and AGN activity will require the inclusion of composite systems
that host AGNs, which entails other measures of AGN luminosity or 
a careful treatment of their spectra in order to separate AGN and SF contributions
to the emission line luminosity. This will be undertaken in future work.

\section{Conclusions}

We investigate the trends between the radio and long-wavelength MIR luminosities of narrow-line (Type II) 
emission-line selected AGNs from the SDSS, which reveals
characteristic patterns in the MIR-Radio plane, suggestive of two distinct populations or ``branches". Building on the
substantial ancillary data from the SDSS, as well as AKARI FIR photometry, we devise various tests to help unveil 
the properties of the two branches. Objects on Branch A are generally radio-quiet, have younger central stellar populations
and exhibit an IR-radio relationship consistent with that of normal inactive SF galaxies, though with suppressed luminosity
at 12 \mics.  Objects on Branch B account for most radio-loud AGN, have evolved central stellar populations and
lie well off the location of SF galaxies on the MIR-Radio plane, implying that their MIR emission is unrelated to SF.
These tests demonstrate that the galaxy-integrated MIR luminosity of AGNs on Branch A arises primarily from dust
heated by on-going SF. A correlation between the \othree\ luminosity and the MIR luminosity suggests a 
possible relationship between SF and nuclear emission, but important selection effects inherent to emission-line selected
AGNs may also account for some or all of the trend. We highlight the result that the majority of Seyfert galaxies
lie on (or are consistent with lying on) Branch A, which suggests a connection between the processes that govern SF 
and the fueling of radiatively-efficient AGN activity. The MIR-Radio plane is a useful tool in studies of the SF
and accretion properties of AGNs over a range of nuclear luminosities.

\acknowledgements

We thank Hagai Netzer and Li Shao for useful discussions.
Funding for the SDSS and SDSS-II has been provided by the Alfred P. Sloan Foundation, the Participating Institutions, the National Science Foundation, the U.S. Department of Energy, the National Aeronautics and Space Administration, the Japanese Monbukagakusho, the Max Planck Society, and the Higher Education Funding Council for England. The SDSS Web Site is http://www.sdss.org/.
This publication makes use of data products from the Wide-field Infrared Survey Explorer, which is a joint project of the University of California, Los Angeles, and the Jet Propulsion Laboratory/California Institute of Technology, funded by the National Aeronautics and Space Administration. This research also employs observations with AKARI, a JAXA project with the participation of ESA.
  
\bibliography{wise_radio}

\end{document}